\documentclass[aps, prd,
twocolumn,%
superscriptaddress,
showpacs, nofootinbib, eqsecnum, amsmath, amssymb, floatfix
]{revtex4}

\usepackage{hyperref}
\usepackage{bm}
\usepackage{graphicx}
\usepackage[usenames]{color}
\usepackage{ulem}
\usepackage{amsmath}
\usepackage{amssymb}

\allowdisplaybreaks

\newcommand{\bmSeffp}{{\bm{S}_{0}^+}}
\newcommand{\bmSeffm}{{\bm{S}_{0}^-}}
\newcommand{\Seffp}{{S_{0}^+}}
\newcommand{\Seffm}{{S_{0}^-}}

\newcommand{\IAP}{\affiliation{Institut d'Astrophysique de Paris,
   UMR 7095 CNRS Universit\'e Pierre \& Marie Curie, 98$^{\text{bis}}$
   boulevard Arago, 75014 Paris, France}}
\newcommand{\Maryland}{\affiliation{Maryland Center for Fundamental
    Physics \& Joint Space-Science Institute,\\ Department of Physics,
University of Maryland, College
    Park, MD 20742, USA}}

\begin{document}

\title{Spin effects on gravitational waves from inspiraling compact binaries
at second post-Newtonian order}

\author{Alessandra Buonanno} \Maryland%
\author{Guillaume Faye} \IAP %
\author{Tanja Hinderer} \Maryland %

\date{\today}

\begin{abstract}
  We calculate the gravitational waveform for spinning, precessing compact
  binary inspirals through second post-Newtonian order in the amplitude. When
  spins are collinear with the orbital angular momentum and the orbits are
  quasi-circular, we further provide explicit expressions for the
  gravitational-wave polarizations and the decomposition into spin-weighted
  spherical-harmonic modes. Knowledge of the second post-Newtonian spin terms
  in the waveform could be used to improve the physical content of analytical
  templates for data analysis of compact binary inspirals and for more
  accurate comparisons with numerical-relativity simulations.
\end{abstract}


\maketitle

\section{Introduction}
\label{sec:intro}

Coalescing compact binary systems are a key source of gravitational
radiation for ground-based gravitational-wave detectors such as the
advanced Laser Interferometer Gravitational Wave Observatory
(LIGO)~~\cite{Abbott:2007}, the advanced Virgo~\cite{Acernese:2008},
the GEO-HF~\cite{Grote:2008zz}, the Large Cryogenic Gravitational
Telescope (LCGT) (or KAGRA)~\cite{Kuroda:2010}, coming into
operation within the next few years, and future space-based
detectors~\cite{lisa,ESALISAwebsite}. For this class of
gravitational-wave sources, the signal detection and interpretation
will be based on the method of matched filtering~
\cite{Finn1992,Finn1993}, where the noisy detector output is cross
correlated with a bank of theoretical templates. The accuracy
requirement on the templates is that they remain as much as possible
phase coherent with the signal over the hundreds to thousands of
cycles of inspiral that are within the detector's sensitive
bandwidth.

Constructing such accurate templates has motivated a significant
research effort during the past 30 years. In the regime where the
separation between the two bodies is large, gravitational waveforms
can be computed using the post-Newtonian (PN) approximation
method~\cite{Sasaki:2003xr,
  Blanchet2006, Futamase:2007zz}. In the post-Newtonian scheme, the results
are written as an asymptotic expansion in powers of $v_A/c$, with
$v_A$ being the magnitude of the orbital coordinate velocity
$\bm{v}_A$ of body $A$ at a given time. This approximation is
physically relevant for $v_A/c \ll 1$, i.e. in the so-called
inspiraling regime where the radiation reaction forces, of order
$\sim (v_A/c)^5 $ are negligible over an orbital period and act
adiabatically on a quasiconservative system. In the domain of
validity of the post-Newtonian scheme, the separation $r \sim (G
m_A/v^2)\sim (c/v)^2$, with $m=m_1 + m_2$ and $v = |\bm{v}| \equiv
|\bm{v}_1 - \bm{v}_2 |$, remains large with respect to the radii of
both compact objects $\sim G m_A/c^2$ or, in other words, the bodies
can be regarded effectively as point particles.

Post-Newtonian waveforms cease to be reliable near the end of the inspiral and
the coalescence phase, where numerical-relativity simulations should be used
to predict the gravitational-wave signal ~\cite{Pretorius2005a,
  Campanelli2006a, Baker2006a}. By combining the information from
post-Newtonian predictions and the numerical-relativity simulations it is
possible to accurately and analytically describe the gravitational-wave signal
during the entire inspiral, plunge, merger and ringdown
process~\cite{Buonanno99, Buonanno00, DJS00, Buonanno-Cook-Pretorius:2007,
  Ajith:2008, Damour2009a, Pan:2009wj, Santamaria:2010yb, Pan:2011gk}.

For nonspinning binaries, the post-Newtonian expansion has been iterated to
$3.5$PN order beyond the leading Newtonian order in the gravitational-wave
phasing~\cite{Blanchet95a, Blanchet04, Blanchet2005b}. The gravitational-wave
amplitude has been computed through $3$PN
order~\cite{Blanchet96a, Kidder07, Kidder2008, BFIS} and the quadrupole mode
through $3.5$PN order~\cite{Faye:2012we}. However, black hole binaries could
potentially have large spins~\cite{Miller2009} which may be misaligned with
the orbital angular momentum, in which case the precession effects add
significant complexity to the emitted gravitational
waves~\cite{Apostolatos1994}. Ignoring the effects of black hole spins could
lead to a reduction in the signal-to-noise ratio and decrease the detection
efficiency~\cite{Apostolatos1996, Buonanno:2002fy} although this should be
overcome with phenomenological and physical
models~\cite{Pan:2003qt, Buonanno2004, Buonanno:2005pt, Buonanno06,
  Ajith:2009, Pan:2009wj, Ajith:2011ec, Brown:2012gs, Taracchini:2012ig}.
To maximize the payoffs for astrophysics will require extracting the source
parameters from the gravitational-wave signal using template models computed
from the most accurate physical prediction
available~\cite{CutlerFlanagan1994, PoissonWill95, vanderSluys, AjithBose2009}.
Spin effects in the waveform are currently known through much lower
post-Newtonian order than for nonspinning binaries. More specifically, spin
effects are known through $2.5$PN order in the
phase~\cite{Mikoczi:2005dn,Faye-Blanchet-Buonanno:2006,
Blanchet-Buonanno-Faye:2006}, $1.5$PN
order in the polarizations for spin-orbit effects~\cite{Kidder:1995zr,
  Arun:2009}, 2PN order for the spin${}_1$-spin${}_2$
effects~\cite{Kidder:1995zr, Will96} and partially $3$PN order in the
polarizations for the tail-induced spin-orbit
effects~\cite{BlanchetEtAl:2011}.

In this paper, we compute all spin effects in the gravitational-wave
strain tensor through 2PN order. This requires knowledge of the
influence of the spins on the system's orbital dynamics as well as
on the radiative multipole moments. At this PN order, nonlinear spin
effects attributable to the spin-induced quadrupole moments of the
compact objects first appear. Using results from
Ref.~\cite{1980AnPhy.130..188B,PortoRothstein2006,Porto:2008jj,SP10},
we derive the stress-energy tensor with self-spin terms and compute
the self-induced quadrupole terms in the equations of motion and in
the source multipole moments at 2PN order. Our results are in
agreement with previous calculations~\cite{Poisson:1997ha,Damour01c,
Steinhoff:2010zz,Porto:2012x}.

The two main inputs entering our calculation of the gravitational-wave strain
tensor through 2PN order are (i) the results of Refs.~\cite{Kidder:1995zr,
  Poisson:1997ha, Blanchet-Buonanno-Faye:2006} for the influence of the spins
on the system's orbital dynamics, which have also been derived by effective
field theory and canonical methods \cite{PortoRothstein2006,Porto:2008tb,
  Porto:2010tr,Damour:2007nc, Steinhoff08, Steinhoff08a, Steinhoff08b}, and (ii) the spin
effects in the system's radiative multipole
moments~\cite{Blanchet-Buonanno-Faye:2006}. Recently, the necessary knowledge
to compute the waveform at 2.5PN order was obtained using the effective field
theory approach \cite{Porto:2010tr,Porto:2012x}. Here we use (i) and (ii) in
the multipolar wave generation formalism~\cite{thorne80, Blanchet:1992br,
  Blanchet:1995fr} to obtain the waveform for spinning, precessing binaries
through 2PN order. To compute the gravitational polarizations from
this result, one must specify an appropriate source frame and
project the strain tensor onto a polarization triad. For precessing
systems, there are several frames that could be
employed~\cite{Finn1993, Kidder:1995zr, Buonanno:2002fy,
Schmidt:2010it,OShaughnessy2011, Ochsner2012,
2011PhRvD..84l4011B,Schmidt:2012rh}. For nonprecessing binaries with
the spins collinear to the orbital angular momentum, the most
natural frame is the one used for nonspinning binaries. Therefore,
instead of choosing one frame, for simplicity, we specialize to the
nonprecessing case and quasicircular orbits and provide the explicit
expressions for the gravitational polarizations. Lengthy
calculations are performed with the help of the scientific software
\textsc{mathematica}{\footnotesize \textregistered}, supplemented by
the package xTensor~\cite{xtensor} dedicated to tensor calculus. Our
generic, precessing result is available in \textsc{mathematica}
format upon request and can be used to compute the polarizations for
specific choices of frame. We notice that the 2PN terms in the
polarizations, for circular orbits, linear in the spins were also
computed in Ref.~\cite{1998PhRvD..57.6168O}. However, these results
contain errors in the multipole moments, which were corrected in
Ref.~\cite{Blanchet-Buonanno-Faye:2006}.

For future work at the interface of analytical and numerical
relativity, we also explicitly compute the decomposition of the
strain tensor into spin-weighted spherical-harmonic modes for
nonprecessing spinning binaries on circular orbits. The
test-particle limit of these results can also be directly compared
with the black-hole perturbation calculations of
Refs.~\cite{Tagoshi:1996gh,Pan2010hz}, and we verify that the
relevant terms agree.

The organization of the paper is as follows. In
Sec.~\ref{sec:modelling}, we review the Lagrangian for compact
objects with self-induced spin effects
~\cite{1980AnPhy.130..188B,PortoRothstein2006,Porto:2008jj,Steinhoff:2010zz},
compute the stress-energy tensor and derive the self-induced spin
couplings in the two-body acceleration and source multipole moments
~\cite{Poisson:1997ha,Damour01c,Steinhoff:2010zz,Porto:2012x}. In
Sec.~\ref{sec:dynamics} we summarize the necessary information about
spin effects in the equations of motion and the wave generation
necessary for our calculation. In Sec.~\ref{sec:SO} we calculate the
spin-orbit effects at 2PN order in the strain tensor for generic
precessing binaries. In Sec.~\ref{sec:SS} we complete the knowledge
of 2PN spin-spin terms by including the spin self-induced quadrupole
terms in addition to the spin${}_1$-spin${}_2$ terms obtained in
Ref.~\cite{Kidder:1995zr}. In Sec.~\ref{sec:pol} we specialize to
quasicircular orbits and explicitly give the polarization tensors
for nonprecessing systems. Then, in Sec.~\ref{sec:modes} we
decompose the polarizations into spin-weighted spherical-harmonic
modes. Finally, Sec.~\ref{sec:conclusions} summarizes our main
findings.

We use lowercase Latin letters $a, b, ..., i, j, ...$ for indices of spatial
tensors. Spatial indices are contracted with the Euclidean metric, with up or
down placement of the indices having no meaning and repeated indices summed
over. We use angular brackets to denote the symmetric, trace-free (STF)
projection of tensors, e.g., $T_{\langle ij\rangle} = {\rm
  STF}[T_{ij}]=T_{(ij)}-\frac{1}{3}\delta_{ij}T_{kk}$, where the round
parentheses indicate the symmetrization operation. Square
parentheses indicate antisymmetrized indices, e.g., $T_{[ij]} =
\frac{1}{2} (T_{ij} - T_{ji})$. The letter $L=i_1... i_\ell$
signifies a multi-index composed of $\ell$ STF indices. The
transverse-traceless (TT) projection operator is denoted ${\cal
P}_{ijab}^\mathrm{TT}={\cal P}_{a(i}{\cal
  P}_{j)b}-\frac{1}{2}{\cal P}_{ij}{\cal P}_{ab}$, where ${\cal
  P}_{ij}=\delta_{ij}-N_iN_j$ is the projector orthogonal to the unit
direction $\bm{N}=\bm{X}/R$ of a radiative coordinate system
$X^\mu=(cT, \bm{X})$, where the boldface denotes a spatial
three-vector. As usual, $g_{\mu\nu}$ represents the space-time
metric and $g$ its determinant. The quantity $\varepsilon_{ijk}$ is
the antisymmetric Levi-Civit\`a symbol, with $\varepsilon_{123}=1$,
and $\epsilon_{\mu\nu\rho\sigma}$ stands for the Levi-Civit\`a
four-volume form, with $\epsilon_{0123} = + \sqrt{-g}$. Henceforth,
we shall indicate the spin${}_1$-spin${}_2$ terms with $S_1S_2$, the
spin${}^2_1$, spin${}^2_2$ terms with $S^2$ and the total spin-spin
terms with ${\rm SS}$. Throughout the paper, we retain only the
terms relevant to our calculations and omit all other terms, which
either are already known or appear at a higher post-Newtonian order
than required for our purposes.

\section{Modeling spinning compact objects with self-induced quadrupoles}
\label{sec:modelling}

In this section we review the construction of a Lagrangian for compact objects
with self-induced quadrupole spin effects
~\cite{Tulczyjew1959,1980AnPhy.130..188B,PortoRothstein2006,Porto:2008jj,
Steinhoff:2010zz},
compute the stress-energy tensor and derive the self-induced spin couplings in
the two-body acceleration and source multipole moments. Our findings are in
agreement with previous
results~\cite{Poisson:1997ha,Damour01c,Steinhoff:2010zz,Porto:2012x}.

\subsection{Lagrangian for compact objects with self-induced spin effects}

A Lagrangian for a system of spinning compact objects with
nondynamical\footnote{We shall not include kinetic terms in the
Lagrangian
  for the quadrupole moment that can describe resonance effects in neutron
  stars.} self-induced quadrupole moments can be obtained by augmenting the
Lagrangian for point particles with $L^{\text{S}^2}_A$ describing
the quadrupole-curvature coupling for each body $A$. Since the
action for body $A$ must admit a covariant representation, the
corresponding Lagrangian $L^{\text{S}^2}_A$ should be a function of
the four-velocity $u_A^\mu$, the metric $g_{\mu\nu}$, the Riemann
tensor $R^\lambda_{~\rho\mu\nu}$ and its covariant derivatives,
evaluated at the worldline point $y_A^\mu$, and the spin variables
entering via the antisymmetric spin tensor $S_A^{\mu\nu}$.

The spin tensor $S_A^{\mu\nu}$ contains six degrees of freedom. It
is well known that in order to reduce them to the three physical
degrees of freedom a spin supplementary condition (SSC) should be
imposed~~\cite{BOC79}. This is equivalent to performing a shift of
the worldline $y_A^\mu$. In this paper we specialize to the SSC
$S_A^{\mu\nu}p^A_\nu=0$ which is equivalent to
$S_A^{\mu\nu}u^A_\nu=0$ since $p_A^\mu \approx m_A c u_A^\mu$
through 2.5PN order. To ensure the preservation of the SSC under the
evolution, we follow Ref.~\cite{Porto:2008jj} and introduce the spin
tensor ${{{{\cal{S}}}}}_A^{\mu\nu} =S_A^{\mu\nu} + 2 u_A^{[\mu}
S_A^{\nu]\lambda} u^A_\lambda$. The spin tensor
${{{{\cal{S}}}}}_A^{\mu\nu}$ automatically satisfies the algebraic
identity ${{{{\cal{S}}}}}^{\mu\nu}_A u^A_\nu = 0$, which provides
three constraints that can be used to reduce the spin degrees of
freedom from six to three.

From the above discussion and Refs.~\cite{Porto:2005ac,
PortoRothstein2006}, we assume that the Lagrangian of particle $A$
is of the form $L^{\text{S}^2}_A=L_{A\, \mu\nu\lambda\rho}
{{{\cal{S}}}}_A^{\mu\nu} {{{\cal{S}}}}_A^{\lambda\rho}$, where
$L_{A\, \mu\nu\lambda\rho}$ is a polynomial in the Riemann tensor
and its derivatives, as well as the four-velocity $u_A^\mu$. As
noticed in Ref.~\cite{DE98a}, any term proportional to
$\nabla_{...}R_{\alpha\beta}$ evaluated at point $y_A^\mu$ can be
recast into a redefinition of the gravitational field. As a result,
the Riemann tensor may be replaced in each of its occurrences by the
Weyl tensor $C^\lambda_{~\rho\mu\nu}$, which can be decomposed into
a combination of the gravitoelectric- and gravitomagnetic-type STF
tidal quadrupole moments $G_{\mu\nu}^A \equiv G_{\mu\nu}(y_A^\alpha)
\equiv - c^2 R_{\mu\alpha\nu\beta} u_A^{\alpha} u_A^{\beta}$ and
$H^A_{\mu\nu}\equiv H_{\mu\nu}(y_A^\alpha) \equiv 2 c^3
R^{A*}_{\mu\alpha\nu\beta} u_A^{\alpha} u_A^{\beta}$ with
$R^*_{\mu\nu\alpha\beta} \equiv \frac{1}{2}
\epsilon_{\mu\nu\rho\sigma} R^{\rho\sigma}_{~~\alpha\beta}$. More
generally, the multiple space derivatives of
$C^\lambda_{~\rho\mu\nu}$ at point $y_A^\mu$ may be expressed in
terms of some STF tidal multipole moments $G^A_{\mu_1 ...\mu_\ell}$
and $H^A_{\mu_1 ... \mu_\ell}$ of parity $1$ and $-1$ respectively.
However, those higher-order moments will play no role in this paper.

Taking into account that the contraction of the velocity vector $u_A^\nu$
with both $G^A_{\mu\nu}$ and ${{{\cal{S}}}}^{\mu\nu}_A$ vanishes, that the
spin and tidal multipole tensors are traceless, and that the Lagrangian must
obey parity and time-reversal
symmetries 
we obtain~\cite{1980AnPhy.130..188B,Porto:2005ac,PortoRothstein2006,
Porto:2008jj}
\begin{equation} \label{eq:LSSA}
L^{\text{S}^2}_A = - \frac{\kappa_A}{2 m_A c^2} G_{\mu\nu} S^\mu_{A
  \lambda}\,S_A^{\lambda \nu} \, .
\end{equation}
Here, we have also assumed that the rotating body is axially symmetric and we
have replaced ${{{\cal{S}}}}^{\mu \nu}_A$ with $S^{\mu \nu}_A$ since the
difference between these spin variables contributes to the equations of motion
at $\mathcal{O}(S_A^3)$, where $S_A= \sqrt{|S_A^\mu S^A_\mu|}$ with $S^A_\mu =
\epsilon_{\rho\sigma\nu\mu} S^{\rho\sigma}_A p^\nu_A/(2 m_A c)$.

For a neutron star the numerical constant $\kappa_A$ in
Eq.~(\ref{eq:LSSA}) depends on the equation of state of the fluid
\cite{Laarakkers99}. For an isolated black hole
$\kappa_A=1$~\cite{Poisson:1997ha,Damour01c}, but for a black hole
in a compact binary $\kappa_A$ can deviate from 1. However, these
deviations occur at PN orders that are much higher than the ones
considered here. We notice that the leading contribution
$\kappa_A=1$ can be obtained by computing the acceleration of body
$A$ from Eq.~(\ref{eq:LSSA}) in a compact binary for $m_A\ll m$ and
matching it with the acceleration of a test particle in the
gravitational field of a Kerr black hole of mass
$m$~\cite{Porto:2005ac}.

\subsection{Effective stress-energy tensor with self-induced quadrupoles}

The piece of the stress-energy tensor encoding the self-induced quadrupole
dynamics of body $A$ reads by definition
\begin{equation} \label{eq:TSSAdef}
T^{\mu\nu}_{\text{quad},A} = \frac{2}{\sqrt{-g}} \frac{\delta}{\delta g_{\mu\nu}(x)}
\int d\tau_A\, L^{\text{S}^2}_A[y_A^\alpha(\tau_A), S^{\alpha\beta}_A(\tau_A)] \, ,
\end{equation}
where $L_A^{\text{S}^2}$ is the Lagrangian~\eqref{eq:LSSA}. To
determine the action of the operator $\delta/\delta g_{\mu\nu}$,
which stands for the usual ``functional derivative'' with respect to
the field $g_{\mu\nu}$, we need to adopt a specific model for the
spin. The rotational state of the extended object $A$ is usually
represented by a tetrad of orthonormal vectors
$e^{\mu}_{A\overline{\alpha}}(\tau_A)$ with $\overline{\alpha} \in
\{0,1,2,3\}$ along the worldline $y^\mu_A$ with affine parameter
$\tau_A$. The corresponding angular rotation tensor is then defined
as $\Omega^{\mu\nu}_A = \eta^{\overline{\alpha}\overline{\beta}}
e_{A\overline{\alpha}}^{\mu} D e^\nu_{A\overline{\beta}}/d\tau_A$.
We now make the reasonable physical hypothesis that the rotation of
the axially symmetric object takes place about the symmetry axis.
The moment of inertia $I_A$ along that direction is a 2PN-order
quantity $ \sim G^2 m_A^3/c^4$ for compactness parameters of order
1, whereas $\Omega^{\mu\nu}_A \sim V_A/R_A$, $R_A$ being the radius
of body $A$ and $V_A$ its typical internal velocity, is roughly
equal to $ c^3/(G m_A)$. In the weak field limit where $G$ goes
formally to zero, the spin must satisfy the relation $S^{\mu\nu}_A =
I_A \Omega^{\mu\nu}_A$, as in special relativity~\cite{HR74}. In the
presence of a nonnegligible gravitational field, this relation is
expected to be modified by nonminimal coupled terms proportional to
positive powers of $R^A_{\mu\nu\alpha\beta}$ times positive powers
of $I_A$ and $S^{\mu\nu}_A$~\cite{Porto:2005ac}:
\begin{equation} \label{eq:hatspin}
\hat{S}_A^{\mu\nu} = I_A
\Big[ \Omega_A^{\mu\nu} + \mathcal{O}\Big(\frac{\hat{S}_A}{c^2}\Big) \Big] \, .
\end{equation}
Here we use a hat to distinguish the generic spin variable from the
one related to our specific spin model. The corrections $ I_A\times
\mathcal{O}(\hat{S}_A/c^2)$ are not relevant for the two-body
dynamics in this paper because they correspond to the 4.5PN order
when taking into account the factor $\mathcal{O}(1/c)$ contained in
the spin variable.

Using the definition~\eqref{eq:hatspin} for the spin variables, we compute in a
covariant manner the variation of the action
\begin{align}
\mathcal{A}^{\text{S}^2} &=
\int d\tau_A L_A^{\text{S}^2}(\tau_A) \nonumber \\ &= \int \frac{d^4
x}{c} \sqrt{-g} \int d\lambda_A L_A^{\text{S}^2}(\lambda_A) \frac{\delta^4(x^\alpha -
y^\alpha_A(\lambda_A))}{\sqrt{-g}} \, ,
\end{align}
when the metric varies by $\delta g_{\mu\nu}(x)$, and find the following
quadrupolar piece of the stress-energy tensor
\begin{align} \label{eq:TSSA}
T^{\mu\nu}_{\text{quad},A} & =  \frac{\kappa_A}{m_A c^2} \Big[ \frac{n^*_A}{2}
\Big(-3 u_A^\mu u^\nu_A G^A_{\lambda \rho} \hat{S}^{\lambda \sigma}_A
\hat{S}_\sigma^{A\rho}\nonumber  \\ & \qquad -  c^2 R^{(\mu}_{A\lambda\rho \tau} u_A^{\nu)}
\hat{S}^\lambda_{A\sigma} \hat{S}_A^{\sigma \rho} u_A^\tau + G^{(\mu}_{A\lambda}
\hat{S}^{\nu)}_{A\rho} \hat{S}_A^{\rho \lambda} \Big) \nonumber \\ & + \nabla_\rho
\Big( I_A c\, n_A^* (G^{A(\mu}_\lambda u_A^{\nu)} \hat{S}_A^{\lambda \rho} -
G_\lambda^{A\rho} \hat{S}^{\lambda(\mu}_A u_A^{\nu)} ) \Big) \Big] \nonumber
\\ & -2 \nabla_\lambda \nabla_\rho
\Big[ n^*_A \hat{S}_A^{\sigma [\lambda} u_A^{(\mu]} \hat{S}_{\sigma}^{A[\nu)}
u_A^{\rho]} \Big] \, ,
\end{align}
where we have indicated with $n^*_A$ the Dirac-type scalar density
$n^*_A(x^\mu) = \int d\lambda_A \, \delta^4(x^\mu-
y_A^\mu(\lambda_A))/\sqrt{-g(x^\nu)}$ and, in the last term, we have adopted
the convention that
symmetrization of indices applies after antisymmetrization.
As derived in Ref.~\cite{Tulczyjew1959}, the most general form of the
effective stress-energy tensor is
\begin{multline}
T^{\mu\nu}_{\text{skel} , A}(x^\mu) = \\\sum_{\ell = 0}^{+\infty}
\nabla_{\lambda_1} \nabla_{\lambda_2} ... \nabla_{\lambda_\ell}
\Big[t_A^{\mu\nu
  | \lambda_1 \lambda_2 ...  \lambda_\ell}(\tau_A) n^*_A(x^\mu) \Big] \, ,
\label{skeletonTmunu}
\end{multline}
where $\tau_A$ is the proper time of the $A$th worldline at event
$y_A^\mu$ with $y_A^0 = x^0$ and the coefficients $t^{\mu\nu |
\lambda_1
  \lambda_2 ... \lambda_\ell}_A(\tau_A)$ are the ``skeleton'' multipole
moments. The latter are not arbitrary but satisfy algebraic
constraints imposed by the equation of conservation $\nabla_\nu
T^{\mu\nu}_\text{skel} =0$. Let us check that we can indeed recast
the total stress-energy tensor, including the monopolar, dipolar and
quadrupolar pieces, in the form (\ref{skeletonTmunu}). If we add
$T^{\mu\nu}_{\text{quad}}$ to the monopolar and dipolar
contributions~\cite{Papa51spin,Tulczyjew1959,Dixon1964,
  Tagoshi-Ohashi-Owen:2001,Faye-Blanchet-Buonanno:2006}
\begin{align} \label{eq:TMon}
T^{\mu\nu}_{\text{mon}+\text{dipole}}  =  \sum_{A} \Big[n_A^*
\tilde{p}_A^{(\mu} u^{\nu)}_A c  + \nabla_\lambda  \Big( n_A^* c\,
u_A^{(\mu} \tilde{S}_A^{\nu) \lambda}\Big)\Big]  \,,
\end{align}
and redefine the spin variable entering the quadrupolar piece as
\begin{equation}
S^{\mu\nu}_A = \tilde{ S}^{\mu\nu}_A  - \frac{2 \kappa_A}{m_A c^2} I_A
\hat{S}_A^{\lambda[\mu} G_{A\lambda}^{\nu]} \,,
\label{S:red}
\end{equation}
we obtain the total stress-energy tensor in the form
\begin{subequations}
\label{eq:TmunuJ}
\begin{align} \label{eq:Tmunu}
T^{\mu\nu} & = \sum_{A} \Big[n_A^* \Big( p_A^{(\mu} u^{\nu)}_A c + \frac{1}{3}
R^{(\mu}_{A\tau\lambda\rho} J_A^{\nu)\tau\lambda\rho}c^2 \Big) \nonumber \\ & \qquad +
\nabla_\lambda  \Big( n_A^*
c\, u_A^{(\mu} S_A^{\nu) \lambda}\Big) \nonumber \\ & \qquad - \frac{2}{3}
\nabla_{\lambda}\nabla_{\rho}\Big(n_A^* c^2 J_A^{\lambda(\mu\nu) \rho}
\Big)\Big] \, ,
\end{align}
where the four-rank tensor $J_A^{\lambda\rho\mu\nu}$ takes the
following expression in our effective description:
\begin{equation}
\label{JA}
J_A^{\lambda\rho\mu\nu}  = \frac{3 \kappa_A}{m_A c^2} S_A^{\sigma[\lambda}
u^{\rho]}_A S^{A[\mu}_\sigma u_A^{\nu]} \, .
\end{equation}
\end{subequations}
Consistently with the approximation already made in the spin
model~\eqref{eq:hatspin}, we have neglected here the difference of
order $I_A\times \mathcal{O}(\hat{S}_A/c^2)$ between the spins
$\hat{S}^{\mu\nu}_A$ and $S^{\mu\nu}_A$ in the above formula. The
net result is that Eq.~(\ref{eq:Tmunu}) matches
Eq.~(\ref{skeletonTmunu}) for $\ell = 0,1,2$ as expected. Moreover,
Eqs.~(\ref{eq:TmunuJ}) agree with Refs.~\cite{Steinhoff:2010zz,SP10}.

Lastly, the conservation of the stress-energy tensor~\eqref{eq:Tmunu} is
equivalent to the equation of motion for the particle worldline, supplemented
by the spin precession equation~\cite{SP10}. They read
\begin{subequations}
\begin{align}  \label{eq:Dixon_EOM}
\frac{D p_A^\mu}{d\tau_A} &= - \frac{c}{2} R^\mu_{A\rho\nu\lambda} u_A^\rho
S^{\nu \lambda}_A - \frac{c^2}{3} \nabla_\tau  R^\mu_{A\rho\nu\lambda}
J_A^{\tau\rho\nu\lambda} \, , \\ \label{eq:Dixon_precession}
\frac{DS^{\mu\nu}}{d\tau_A} &= 2 c \, p_A^{[\mu} u_A^{\nu]} + \frac{4c^2}{3}
R^{[\mu}_{A\tau\lambda\rho} J_A^{\nu]\tau\lambda\rho} \, .
\end{align}
\end{subequations}
Those equations are in full agreement with the equations of evolution derived
from the Dixon formalism truncated at the quadrupolar order~\cite{Dixon1974}.

\subsection{Self-induced quadrupole terms in the 2PN binary dynamics and
source multipole moments}

Once the stress-energy tensor has been derived, the post-Newtonian equations
of motion and the source multipole moments parametrizing the linearized
gravitational field outside the system can be computed by means of the usual
standard techniques~\cite{Blanchet2006}. At 2PN order, the accelerations
including the self-spin interactions were obtained in
Refs.~\cite{Poisson:1997ha,Damour01c}, but the self-induced quadrupole effects
in the source multipole moments were never explicitly included in the standard
version of the post-Newtonian scheme, although recently they were calculated
at 3PN order using effective-field-theory techniques~\cite{Porto:2010zg}. Here
we can use the results of the previous section, which constitutes a natural
extension of the standard post-Newtonian approximation for spinning compact
bodies~\cite{Faye-Blanchet-Buonanno:2006}, and explicitly derive the
self-induced quadrupole couplings in the 2PN dynamics and source multipole
moments.

Henceforth, we define the spin vectors $S_A^i$ by the relation
$S_i^A/c = g_{ij}^A S_A^j$, where $S_i^A$ is the three-form induced
on the hypersurface $t= {\rm const}$ by $S_\mu^A$. Note that it is
$S^i_A/c$ that has the dimension of a spin, while $S^i_A$ has been
rescaled in order to have a nonzero Newtonian limit for compact
objects.

In the post-Newtonian formalism for point particles in the harmonic gauge, it
is convenient to represent effectively the source by the mass density $\sigma
= (T^{00}+T^{ii})/c^2$, the current density $\sigma_i = T^{0i}/c$, and the
stress density $\sigma_{ij}= T^{ij}$. They are essentially the components of
the stress-energy tensor rescaled so as not to vanish in the formal limit $c
\to 0$ for weakly stressed, standard matter. At 2PN order, the second term in
the right-hand side of Eq.~(\ref{eq:Tmunu}) does not contribute. From the
last term, we obtain the following self-spin
contributions:
\begin{subequations}
\begin{align} \label{eq:sigma}
& \sigma^{\text{S}^2} = \frac{\kappa_1}{2 m_1 c^2} \partial_{ij} [\delta_1
S_1^{ki} S_1^{kj}]
+ 1 \leftrightarrow 2 + \mathcal{O}\Big(\frac{S_A^2}{c^4} \Big) \, ,\\
& \sigma^{\text{S}^2}_i =  \mathcal{O}\Big(\frac{S_A^2}{c^2}\Big) \, , \\
& \sigma^{\text{S}^2}_{ij} =  \mathcal{O}\Big(\frac{S_A^2}{c^2}\Big) \, .
\end{align}
\end{subequations}
where $1 \leftrightarrow 2$ represents the counterpart of the
preceding term with particles 1 and 2 exchanged, and $\delta_1
\equiv \delta^3(\bm{x}-\bm{y}_1)$.

At 2PN order, the spin$^{2}$ part of the equations of
motion~\eqref{eq:Dixon_EOM} for, say, the first particle, reduce to
\begin{equation}
\frac{D(u_i^1 c)}{d\tau_1} = \text{non-}S_1^2\text{ terms}
-\frac{\kappa_1}{2m_1^2} \partial_k R^1_{i0j0} S_1^{lk} S_1^{lj} + \mathcal{O}
\Big(\frac{S_1^2}{c^4} \Big) \, .
\end{equation}
The only occurrence of self-spin interactions at 2PN order on the
left-hand side of the above equation comes from the gradient of the
time component of the metric, $g_{00} = -1 + 2 V/c^2 +
\mathcal{O}(1/c^4)$, where the Newton-like potential $V$ satisfies
$\Box V = - 4\pi G \sigma$. Although $V$ coincides with the
Newtonian potential $U$ in the leading approximation, it contains
higher order corrections, including quadratic-in-spin terms coming
from the mass density~\eqref{eq:sigma}, which are smaller than $U$
by a factor $\mathcal{O}(1/c^4)$. They read
\begin{align}
V_{\text{S}^2}  &= - \frac{2\pi G \kappa_1}{m_1 c^2} \partial_{ij} \Delta^{-1}
[\delta_1 S_1^{ki} S_1^{kj}] + 1 \leftrightarrow 2 + \mathcal{O}
\Big(\frac{S_A^2}{c^4} \Big) \nonumber \\
& =  \frac{ G \kappa_1}{2m_1 c^2} \partial_{ij} \frac{1}{r_1} S_1^{ki} S_1^{kj} + 1
\leftrightarrow 2 + \mathcal{O}  \Big(\frac{S_A^2}{c^4} \Big)\, ,
\end{align}
with $\partial_i = \partial/\partial x^i$ and $r_1\equiv |\bm{x}-\bm{y}_1|$,
the symbol $\Delta^{-1}$ holding for the retarded integral operator. Other
potentials appear at the 1PN approximation or beyond, but their sources cannot
contain a self-induced quadrupole below $\mathcal{O}(1/c^4)$; thus they are
negligible here. The self-induced spin part of the acceleration $\bm{a}_1$ of
the first particle is therefore given by
\begin{equation}
  (a^i_1)_{\text{S}^2} = -c^2 (\Gamma_{~0 i}^0)_{\text{S}^2} -
  \frac{\kappa_1}{2m_1^2} \partial_k R^1_{i0j0} S_1^{lk} S_1^{lj} +
  \mathcal{O}  \Big(\frac{S_A^2}{c^4} \Big) \ .
\end{equation}
Replacement of the Christoffel symbols $\Gamma_{~\mu\nu}^\lambda$ and the Riemann
tensor by the leading order values
\begin{align}
\Gamma_{~0 i}^0 = - \frac{\partial_i V}{c^2} +
\mathcal{O}\Big(\frac{1}{c^4}\Big) \, , ~
R_{i0j0} = - \frac{ \partial_{ij} U}{c^2} +
\mathcal{O}\Big(\frac{1}{c^4}\Big) \, ,
\end{align}
with $U = G m_1/r_1 + G m_2 /r_2 + \mathcal{O}(1/c^2)$ yields the more
explicit result (posing $\partial_{1i} \equiv \partial/\partial y_1^i$):
\begin{align}
\label{aSS}
(a^i_1)_{\text{S}^2} &= - \frac{G}{2c^2} \partial_{1ijk} \frac{1}{r} \Big[
\frac{\kappa_2}{m_2} S_2^j S_2^k + \frac{m_2 \kappa_1}{m_1^2} S_1^j S_1^k
\Big] \nonumber \\ &+ \mathcal{O} \Big( \frac{1}{c^4} \Big) \, ,
\end{align}
which agrees with Refs.~\cite{Poisson:1997ha,Damour01c} in the
center-of-mass frame, for $S_A^i/c = \varepsilon_{ijk} S^{jk} +
\mathcal{O}( 1/c^3)$.

Self-induced quadrupolar deformations of the bodies also produce 2PN-order
terms in the source multipole moments $I_L$ and $J_L$. Those are defined as
volume integrals whose integrands are certain polynomials in the densities
$\sigma$, $\sigma_i$ and $\sigma_{ij}$ as well as some gravitational
potentials, such as $V$, that parametrize the metric. Now, since those
potentials are multiplied by prefactors of order $\mathcal{O}(1/c^2)$ and
cannot contain themselves spin${}^2$ interactions below the 2PN order,
monomials involving one potential or more may be ignored for the calculation.
The remaining sources are linear in the $\sigma$ variables. With the help of
the general formula~(5.15) of Ref.~\cite{Blanchet98}, it is then immediate to
get the self-spin contribution to $I_L$:
\begin{equation}
I_L^{\text{S}^2} =\int d^3\!\bm{x} \, x^{\langle i_1}\!\! ...\, x^{i_\ell
  \rangle}\sigma_{\text{S}^2}  +
\mathcal{O} \Big(\frac{S_A^2}{c^4} \Big) \, .
\end{equation}
Inserting expression~\eqref{eq:sigma} for $\sigma_{\text{S}^2}$ and
performing a straightforward integration, we arrive at
\begin{equation}
I_L^{\text{S}^2} = \frac{\kappa_1}{2m_1 c^2} \partial_{1ij}
(y_1^{\langle i_1}\! \! ...\, y_1^{i_\ell \rangle}) S_1^{ki} S_1^{kj} +  1
\leftrightarrow 2 + \mathcal{O} \Big( \frac{S_A^2}{c^4}\Big) \, .
\end{equation}
We can show similarly that $J_L$ is of order $\mathcal{O}(S_A^2/c^2)$. As a
result, at the accuracy level required for the 2PN waveform, the only terms
quadratic in one of the spins that originate from the source moments come
from the quadrupole $\ell=2$, for which we have
\begin{equation}
\label{eq:ISSresult}
I_{ij}^{\text{S}^2} = - \frac{\kappa_1}{m_1 c^4} S_1^{\langle i} S_1^{j\rangle} + 1
\leftrightarrow 2 + \mathcal{O} \Big( \frac{1}{c^6}\Big)\, ,
\end{equation}
whereas similar terms in $(I_L)_{\ell \ge 3}$ or $(J_L)_{\ell \ge
2}$ lie beyond our approximation. The above correction to the mass
quadrupole agrees with that of Porto \textit{et
al}.~\cite{Porto:2010zg} truncated at 2PN order. It is formally of
order $\mathcal{O}(1/c^4)$ but, because $\dot{\bm{S}}_A=
\mathcal{O}(1/c^2)$, it is cast to the 3PN order in the waveform
expansion given below [see Eq.~\eqref{eq:hij}] after the second time
derivative is applied. This result was already argued in
Ref.~\cite{Racine2008}.

\section{Two-body dynamics with spin effects through 2PN order}
\label{sec:dynamics}

The equations of motion in harmonic coordinates for the relative orbital
separation $\bm{x}=r\,\bm{n}$ in the center of mass frame are
~\cite{Blanchet2006}
\begin{subequations}
 \label{eq:eom}
\begin{eqnarray}
\frac{d^2 x^i}{dt^2}&=&a^i_{\rm Newt}+\frac{1}{c^2}a^i_{\rm 1PN}+
\frac{1}{c^3}a^i_{\rm SO} \nonumber \\
&& +\frac{1}{c^4}\left[a^i_{\rm S_1S_2}+a^i_{\text{S}^2}+
a^i_{\rm 2PN}\right], \label{eq:eomscale}
\end{eqnarray}
where
\begin{align}
\bm{a}_{\rm Newt}&=-\frac{G m}{r^2}\bm{n} \, , \\
\bm{a}_{\rm 1PN}&=-\frac{G m}{r^2}\left\{\left[(1+3\nu)v^2-
\frac{3}{2}\nu\dot r^2-2(2+\nu)\frac{G m}{r}\right]\bm{n}  \right.\nonumber\\
& \qquad  \qquad  -2 \dot r (2-\nu) \bm{v}\Bigg\},
\end{align}
with $m\equiv m_1+m_2$, $\nu\equiv m_1\,m_2/m^2$, $\bm{n}=\bm{x}/r$ and
$\bm{v}=d\bm{x}/dt$.
The 2PN acceleration given, e.g., in Ref.~\cite{Kidder:1995zr} will not be
needed for our calculation. The spin-orbit terms are~\cite{Kidder:1995zr}
\begin{align}
&\bm{a}_{\rm SO}=\frac{G}{r^3} \left\{6\left[(\bm{n} \times \bm{v})\cdot
    \left(2 \bm{S} +
\delta\, \bm{\Sigma}\right)\right]\bm{n} \right. \label{eq:aSO} \\
& \left. \qquad \qquad - \left[\bm{v}\times \left(7 \bm{S}+
3\delta\, \bm{\Sigma}\right)\right] +3\dot r \left[\bm{n}\times \left(3 \bm{S}+
\delta\,\bm{\Sigma}\right)\right] \right\} \, ,\nonumber
\end{align}
\end{subequations}
where we denote with $\delta=(m_1-m_2)/m$ and
\begin{subequations}\label{SDelta}\begin{align}
\bm{S} &\equiv \bm{S}_1 + \bm{S}_2\,,\\ \bm{\Sigma}
&\equiv m\left[\frac{\bm{S}_2}{m_2} -
\frac{\bm{S}_1}{m_1}\right]\,.
\end{align}\end{subequations}
The spin$_1$-spin$_2$ interaction terms are~\cite{Kidder:1995zr}
\begin{subequations}
\begin{align}
&\bm{a}_{\rm S_1 S_2}=-\frac{3G}{m \nu r^4}\bigg[ \left[
( \bm{S}_1 \cdot \bm{S}_2)-5 (\bm{n}\cdot \bm{S}_1)
(\bm{n} \cdot \bm{S}_2)\right]\bm{n}    \nonumber\\
&  \qquad \qquad \qquad \quad +(\bm{n} \cdot \bm{S}_1)\bm{S}_2+(\bm{n}\cdot
\bm{S}_2)\bm{S}_1 \bigg].
\end{align}
As originally computed in Ref.~\cite{Poisson:1997ha} [see
Eq.~(\ref{aSS}) above], an additional term due to the influence of
the spin-induced mass quadrupole moment on the motion arises at 2PN
order:
\begin{align}
&\bm{a}_{\text{S}^2}=-\frac{3G}{2m \nu r^4}\bigg\{
\bm{n}\left[\frac{\kappa_1}{q} S_1^2+q \ \kappa_2 S_2^2 \right]\nonumber\\
& \ \ \ +2 \left[\frac{\kappa_1}{q}(\bm{n} \cdot \bm{S}_1)\bm{S}_1 +
q \, \kappa_2 (\bm{n}\cdot \bm{S}_2)\bm{S}_2\right]\nonumber\\
& \ \ \ -\bm{n}\left[\frac{5\kappa_1}{q} (\bm{n}\cdot \bm{S}_1)^2+5 q \,
\kappa_2 (\bm{n}\cdot \bm{S}_2)^2\right]\bigg\}. \label{eq:aspinspin}
\end{align}
\end{subequations}
Here, $q=m_1/m_2$ is the mass ratio and we recall that the parameters
$\kappa_A$ characterize the mass quadrupole moments of the
bodies.

We find that the quadratic spin contribution to the acceleration can
be rewritten in a simpler way by introducing the spin variables
\begin{align}
\bmSeffp &= \frac{m}{m_1} \left(\frac{\kappa_1}{\kappa_2}
\right)^{1/4} (1+ \sqrt{1-\kappa_1 \kappa_2})^{1/2} \bm{S}_1 \nonumber \\ &+
\frac{m}{m_2} \left(\frac{\kappa_2}{\kappa_1}
\right)^{1/4} (1- \sqrt{1-\kappa_1 \kappa_2})^{1/2} \bm{S}_2 \, ,
\end{align}
and $\bmSeffm$, which is obtained by exchanging the labels 1 and 2
in the above equation.~\footnote{In the formal limit where the
induced quadrupole of
  at least one body vanishes, so that e.g. $\kappa_2 \to 0$, we may define the
  effective spins as: $\bmSeffp = \frac{m}{m_1} \sqrt{2} \bm{S}_1$, $\bmSeffm
  = \frac{m}{m_1} \frac{\kappa_1}{\sqrt{2}} \bm{S}_1 + \frac{m}{m_2} \sqrt{2}
  \bm{S}_2$.} Those variables generalize the quantity $\bm{S}_0$ of
Ref.~\cite{Damour01c} in the case where the two bodies are not black
holes. In terms of these spin variables the spin-spin part of the
acceleration reads
\begin{multline}
\label{eq:ssaccel}
\bm{a}_{\rm S_1 S_2} + \bm{a}_{\text{S}^2} = - \frac{3G}{2 m r^4} [ \bm{n} \,
(\bmSeffp \cdot \bmSeffm) + (\bm{n} \cdot
\bmSeffp) \, \bmSeffm
\\   +  (\bm{n} \cdot \bmSeffm) \, \bmSeffp -
5  \bm{n} \, (\bm{n} \cdot \bmSeffp)  (\bm{n} \cdot
\bmSeffm) ]
\, .
\end{multline}

The spin precession equations through 2PN order are~\cite{Kidder:1995zr,
  Racine:2008qv}
\begin{subequations}
\begin{align}
&\frac{d\bm{S}}{dt}=\frac{G m \nu}{c^2 r^2}\bigg\{
\left[-4 (\bm{v}\cdot \bm{S})-
2 \delta\,(\bm{v}\cdot \bm{\Sigma})\right]\bm{n} \nonumber\\
& \qquad~ +\left[3 (\bm{n}\cdot \bm{S})+
\delta\,(\bm{n}\cdot \bm{\Sigma})\right]\bm{v}+\dot r \left[2 \bm{S}+
\delta\,\bm{\Sigma}\right]\bigg\},\\
&\frac{d\bm{\Sigma}}{dt}= \frac{G m}{c^2 r^2}\bigg\{
\left[-2 \delta\,(\bm{v}\cdot \bm{S})-
2(1-2\nu)(\bm{v}\cdot \bm{\Sigma})\right]\bm{n}\nonumber\\
& \qquad~ +\left[\delta\, (\bm{n}\cdot \bm{S})+
(1-\nu)(\bm{n}\cdot \bm{\Sigma})\right]\bm{v}\nonumber\\
& \qquad~ +  \dot r \left[\delta\,\bm{S}+
(1-2\nu)\bm{\Sigma}\right]\bigg\}.
\end{align}
\end{subequations}

It is often convenient to use a different set of spin variables
$S^{\rm c}_{\!Ai}$ whose magnitude remains constant and that obey
precession equations of the form $d{\bm{S}}^{\rm
c}_A/dt=\bm{\Omega}_A \times \bm{S}_A^{\rm c}$. The relationship
between the spin variables appearing in the equations of motion
above and the constant magnitude spin variables is~\cite{
Blanchet-Buonanno-Faye:2006}
\begin{subequations}
\begin{align}
\bm{S}_{\rm c}&= \bm{S}+\frac{G m \nu}{r c^2}\left[2 \bm{S}+
\delta\,\bm{\Sigma}\right]\nonumber\\
&-\frac{\nu}{2 c^2}\left[(\bm{v}\cdot \bm{S})+
\delta\,(\bm{v}\cdot \bm{\Sigma})\right]\bm{v}\, ,\\
\bm{\Sigma}_{\rm c}&= \bm{\Sigma}+\frac{G m}{r c^2}\left[\delta\,\bm{S}
+(1-2\nu)\bm{\Sigma}\right]\nonumber\\
&-\frac{1}{2 c^2}\left[\delta\,(\bm{v}\cdot \bm{S})+
(1-3\nu)(\bm{v}\cdot \bm{\Sigma})\right]\bm{v}. \ \ \ \ \ \ \ \
\end{align}
\end{subequations}

\section{Waveforms with spin effects at 2PN order }

\subsection{General formalism}
\label{subsec:general}

The gravitational radiation from the two-body system is calculated from
symmetric trace-free radiative multipole moments $I_{L}$ and $J_{L}$ using the
general formula from Ref.~\cite{thorne80} truncated at 2PN order
\begin{align} \label{eq:hij}
&h_{ij}^{\mathrm{ TT}}=\frac{2G}{Rc^4}\bigg\{I_{ab}^{(2)}+
\frac{1}{3c} I_{ abc}^{(3)}N^c+\frac{1}{12 c^2}I_{abcd}^{(4)}
N^{c}N^{d}\nonumber\\ &+
\frac{1}{60 c^3}I^{(5)}_{ abcde}N^{c}N^{d}N^{e}+\frac{1}{360c^4}
I^{(6)}_{abcdef}N^{c}N^{d}N^{e}N^{f} \nonumber\\
& + N^{k}\varepsilon_{cka}\bigg[\frac{4}{3c}J_{ bc}^{(2)}+
\frac{1}{2c^2}J_{ bcd }^{(3)}N^d +\frac{2}{15 c^3}
J_{ bcde}^{(4)}N^{d}N^{e}\nonumber\\
& \qquad \qquad +
\frac{1}{36c^4}J_{ bcdef}^{(5)}N^{d}N^{e}N^{f}\bigg]\bigg\}{\cal P}_{ijab}^{\mathrm{TT}},
\end{align}
where $\bm{N}$ is the unit vector pointing from the center of mass
of the source to the observer's location and $R$ is the distance
between the source and the observer. Here, the superscript ${(n)}$
signifies the $n$th time derivative, and the transverse-traceless
projection operator is
\begin{equation}
{\cal P}_{ijab}^{\mathrm{TT}}={\cal P}_{a(i}{\cal P}_{j)b}-
\frac{1}{2}{\cal P}_{ij}{\cal P}_{ab},
\end{equation}
with ${\cal P}_{ij}=\delta_{ij}-N_{i}N_{j}$.

The gravitational radiation (\ref{eq:hij}) can be rewritten in a
post-Newtonian expansion as
\begin{align}
\label{eq:hexpansion} h_{ij}^{\mathrm{TT}} =&\frac{1}{c^4}\,\bigg[
h^{\rm Newt}_{{ij~{\mathrm{TT}}}} + \frac{1}{c^2}\,h^{\rm
1PN}_{{ij~{\mathrm{TT}}}}+\frac{1}{c^2}\,h^{\rm 1PN SO}_{{ij \
{\mathrm{TT}}}} +
\frac{1}{c^3}\,h^{\rm 1.5PN  SO}_{{ij \   {\mathrm{TT}}}} \hfill ~ \; \; \; \;  \; ~ \; \nonumber \\
& \; \; + \frac{1}{c^4}\,h^{\rm 2PN}_{{ij \   {\mathrm{TT}}}} +
\frac{1}{c^4}\,h^{\rm 2PN  SO}_{{ij \   {\mathrm{TT}}}} +
\frac{1}{c^4}\,h^{\rm 2PN  SS}_{{ij \   {\mathrm{TT}}}} + \cdots\bigg]\,.  
\end{align}
The 1PN and 1.5PN spin terms are given explicitly in Refs.
\cite{Kidder:1995zr, Arun:2009}. The terms in the source multipole
moments that are \textit{a priori} needed to compute the spin-orbit
waveform exactly at 2PN order are identified by considering their
schematic structure,
\begin{subequations}
\begin{align} I_{L} &=I_{L}^{\rm Newt}+
\frac{1}{c^2}I_{L}^{\rm 1PN}+\frac{1}{c^3}I_{L}^{\rm SO}\nonumber \\ & \qquad
\quad~~ +
\frac{1}{c^4}(I_{L}^{\rm 2PN}+I_L^{\rm SS}) \,, \label{eq:ulstruct}\\
J_{L}&=J_{L}^{\rm Newt}+\frac{1}{c}J_{L}^{\rm
SO}+\frac{1}{c^2}J_{L}^{\rm 1PN} \nonumber \\ & \qquad
\quad~~ +
\frac{1}{c^3}J_{L}^{\rm 1.5PNSO}\,,
\label{eq:vlstruct}
\end{align}
\end{subequations}
together with the scalings of Eqs.~(\ref{eq:hij}) and
(\ref{eq:eomscale}). Specifically, the following pieces are
required: $(I_{abc}^{\rm Newt})^{(3)}$ using the $1.5$PN motion and
$(I_{abc}^{\rm SO})^{(3)}$ with $\bm{a}^{\rm Newt}$, $(J_{ab}^{\rm
SO})^{(2)}$ with the 1PN motion and the spin evolution, $(J_{ab}^{
\rm 1.5PNSO})^{(2)}$ with $\bm{a}^{\rm Newt}$, $(J_{ab}^{ \rm
Newt})^{(2)}$ with the $1.5$PN accurate motion, and $(J_{abcd}^{\rm
SO})^{(4)}$ with $\bm{a}^{\rm Newt}$. For the SS part, we need
$(I_{ab}^{\rm Newt})^{(2)}$ with $a^{\rm SS}$, as the time
derivative of $I_{ab}^{\rm SS}$ does not contribute at 2PN order.
When we write the waveform in terms of the constant magnitude spin
variables, there is an additional contribution to the 2PN spin piece
of the waveform coming from $J_{ab}^{\rm SO}$ with $\bm{a}^{\rm
Newt}$ and the 1PN conversion factor in $\Sigma^{\rm c}$. The
relevant spin contributions to the multipole moments
are~\cite{Blanchet-Buonanno-Faye:2006}
\begin{widetext}
\begin{subequations}
\begin{align}\label{JijS}
J^{\rm spin}_{ij} &=
\frac{\nu}{c}\biggl\{-\frac{3}{2} r \,n^{\langle
i}\, \Sigma^{ j\rangle}\biggr\}\nonumber\\
&+
\frac{\nu}{c^3}\biggl\{\left(\frac{3}{7}-\frac{16}{7}\nu\right)
r \, \dot r\,v^{\langle i}\, \Sigma^{j\rangle} +
\frac{3}{7} \,\delta \, r \, \dot r
\,v^{\langle i}\,S^{j\rangle} +
\left[\left(\frac{27}{14}-\frac{109}{14}\nu\right)
(\bm{v}\cdot\bm{\Sigma}) + \frac{27}{14} \delta
\,(\bm{v}\cdot\bm{S})\right] r \, n^{\langle
i}\, v^{j\rangle}\nonumber\\
& \qquad  +
\left[\left(-\frac{11}{14}+\frac{47}{14}\nu\right)
(\bm{n}\cdot\bm{\Sigma}) - \frac{11}{14} \delta
\,(\bm{n}\cdot\bm{S})\right]r \,  v^{\langle i} \ v^{j\rangle} +
\left[\left(\frac{19}{28}+\frac{13}{28}\nu\right) \frac{G m}{r} +
\left(-\frac{29}{28}+\frac{143}{28}\nu\right) v^2\right] r \, n^{\langle
i} \, \Sigma^{ j\rangle}\nonumber\\
&\qquad  +
\left[\left(-\frac{4}{7}+\frac{31}{14}\nu\right) (\bm{n}\cdot
\bm{\Sigma}) - \frac{29}{14} \delta
\,(\bm{n}\cdot \bm{S})\right] G m  \, n^{\langle
i}\, n^{j\rangle} + \left[-\frac{1}{14}\frac{G m}{r} -
\frac{2}{7} v^2\right] \delta\, r
\, n^{\langle i}\, S^{j\rangle}\biggr\}\,, \\
I^{\rm spin}_{ijk} &=
\frac{\nu}{c^3} \, r^2 \, \biggl\{-\frac{9}{2}\,\delta
\,n^{\langle i}n^j(\bm{v}\times\bm{S})^{k\rangle}-
\frac{3}{2}\,(3-11\nu)\,n^{\langle i}n^j(\bm{v}\times\bm{\Sigma})^{k\rangle}
\nonumber\\
& \qquad \quad~~ +3\,\delta \,n^{\langle i} v^j(\bm{n}\times\bm{S})^{k\rangle}+
3\,(1-3\nu)\,n^{\langle i}v^j(\bm{n}\times \bm{\Sigma})^{k\rangle}\biggr\} \,,\\
J^{\rm spin}_{ijkl} &=
-\frac{5\nu}{2 c} \, r^3 \, \left\{\delta\, n^{\langle i}n^j n^k S^{l\rangle}+
(1-3\nu)n^{\langle i}n^{j} n^{k}\Sigma^{l\rangle}\right\}\,.
\end{align}
\label{spinmultis}
\end{subequations}
\end{widetext}

The nonspinning contributions to the multipole moments that we employed in our
calculation are
\begin{subequations}
 \label{nonspinmultis}
\begin{align}
I_{ij} &=m\nu  \, r^2 \, n^{\langle i}n^{j\rangle}\,, \\
I_{ijk} &= -m\nu\, r^3 \, \delta  \, n^{\langle i} n^j n^{k \rangle}\,, \\
J_{ij} &=- m\nu\, r^2 \,  \delta \,
\varepsilon_{ab\langle i} n^{j\rangle}n^{ a}v^b \,.
\end{align}
\end{subequations}

\subsection{Spin-orbit effects}
\label{sec:SO}
Using the multipole moments of Eqs. (\ref{spinmultis}) and
(\ref{nonspinmultis}) in Eq. (\ref{eq:hij}) and substituting the equations of
motion (\ref{eq:eom}) and (\ref{eq:aspinspin}), we find the following 2PN
spin-orbit piece:
\begin{widetext}
\begin{align}
\label{eq:hijSO}
h_{ij \ {\rm{TT}}}^{\rm 2PN SO}&=
\frac{2 G^2 m\nu}{r^2R}{\cal P}^{\rm TT}_{ijab}\Bigg\{
n^a \, n^b \left[\frac{5}{2} (3-13\nu) \, \dot r^2 \, (\bm{n}\times
\bm{\Sigma}_{\rm c})\cdot \bm{N} +30 (1-4\nu)(\bm{n}\cdot \bm{N}) \,
\dot r \, (\bm{n}\times\bm{v})\cdot \bm{\Sigma}_{\rm c}
\right.\nonumber\\
& \left. \; \; \; -
(7-29\nu) \, \dot r \, (\bm{v}\times \bm{\Sigma}_{\rm c})\cdot \bm{N}
 -6(1-4\nu) (\bm{v}\cdot
\bm{N})(\bm{n}\times\bm{v})\cdot \bm{\Sigma}_{\rm c} -
\frac{1}{2}(3-13\nu) \, v^2
\, (\bm{n}\times \bm{\Sigma}_{\rm c})\cdot \bm{N}
\right.\nonumber\\
& \left. \; \; \; - \frac{2G m}{3r}(1-5\nu)
(\bm{n}\times \bm{\Sigma}_{\rm c})\cdot \bm{N} +\delta \left(
\frac{35}{2}\,  \dot r^2 \, (\bm{n}\times \bm{S}_{\rm c})\cdot \bm{N}-
\frac{7}{2} \, v^2 \,  (\bm{n}\times \bm{S}_{\rm c})\cdot \bm{N} +
\ 60 (\bm{n}\cdot \bm{N}) \, \dot r (\bm{n}\times\bm{v})\cdot \bm{S}_{\rm c}
\right.\right. \nonumber\\
&  \; \; \;  -
12(\bm{v}\cdot \bm{N})(\bm{n}\times\bm{v})\cdot \bm{S}_{\rm c} -13 \, \dot r \,
(\bm{v}\times \bm{S}_{\rm c})\cdot \bm{N} \bigg) \bigg]\, +
\, n^{a}(\bm{n}\times \bm{S}_{\rm c})^{b} \delta \bigg[
35 (\bm{n}\cdot \bm{N}) \, \dot r^2-
14 (\bm{v}\cdot \bm{N}) \, \dot r-
7 (\bm{n}\cdot \bm{N}) \, v^2 \bigg]\nonumber\\
& + n^{a}(\bm{n}\times \bm{N})^{b}\left[\frac{5}{2}(3-13\nu)
\, \dot r^2 \, (\bm{n}\cdot \bm{\Sigma}_{\rm c})-\frac{1}{2}(3-13\nu) \,
v^2  \, (\bm{n}\cdot \bm{\Sigma}_{\rm c})+
\frac{15}{2}(1-3\nu) \, \dot r^2
\, (\bm{n}\cdot \bm{N}) (\bm{N}\cdot \bm{\Sigma}_{\rm c}) \right.\nonumber\\
& \left. \; \; \;   -5 (1-3\nu)\, \dot r \,
(\bm{v}\cdot \bm{N}) (\bm{N}\cdot \bm{\Sigma}_{\rm c}) -
\frac{3}{2}(1-3\nu) \, v^2
\, (\bm{n}\cdot \bm{N})(\bm{N}\cdot \bm{\Sigma}_{\rm c})-\frac{2Gm}{r}(1-3\nu)
(\bm{n}\cdot \bm{N}) (\bm{N}\cdot \bm{\Sigma}_{\rm c})\right.\nonumber\\
&\left. \; \; \; +
\frac{4Gm}{3r}(1-5\nu) (\bm{n}\cdot \bm{\Sigma}_{\rm c}) -
(3+11\nu) \, \dot r \,  (\bm{v}\cdot \bm{\Sigma}_{\rm c})+
\delta \left(\frac{4Gm}{r} (\bm{n}\cdot \bm{S}_{\rm c}) +
\frac{35}{2}\, \dot r^2 \, (\bm{n}\cdot \bm{S}_{\rm c}) -
\frac{7}{2}  \, v^2  \, (\bm{n}\cdot \bm{S}_{\rm c})\right. \right.\nonumber\\
&\left. \; \; \;+
\frac{15}{2}  \, \dot r^2 \,(\bm{n}\cdot \bm{N}) (\bm{N}\cdot \bm{S}_{\rm c})-
\frac{2Gm}{r}(\bm{n}\cdot \bm{N}) (\bm{N}\cdot \bm{S}_{\rm c})-
\frac{3}{2}  \, v^2  \, (\bm{n}\cdot \bm{N})(\bm{N}\cdot \bm{S}_{\rm c}) -
5 \, \dot r \, (\bm{v}\cdot \bm{N}) (\bm{N}\cdot \bm{S}_{\rm c}) +
\, \dot r \, (\bm{v}\cdot \bm{S}_{\rm c}) \bigg) \right]
\nonumber\\
&+ n^{a}(\bm{n}\times \bm{\Sigma}_{\rm c})^{b}
\bigg[5 (3-13\nu)(\bm{n}\cdot \bm{N})\, \dot r^2\, -
(3-13\nu) (\bm{n}\cdot \bm{N})\, v^2-2(3-14 \nu)(\bm{v}\cdot \bm{N}) \,
\dot r  \nonumber\\
&\left. \; \; \; -
\frac{4Gm}{3r}(1-5\nu)(\bm{n}\cdot \bm{N}) \right]\, +
\, n^{a}(\bm{n}\times\bm{v})^{b} \, \dot r \,  \left[2
(1-4\nu)(\bm{N}\cdot \bm{\Sigma}_{\rm c})+6 \delta\,
(\bm{N}\cdot \bm{S}_{\rm c})\right]\nonumber\\
&+ (\bm{n}\times\bm{N})^{a} \Sigma_{\rm c}^{b}
\left[ \frac{5}{4} (1+7\nu)\, \dot r^2 +
\frac{15}{4} (1-3\nu) (\bm{n}\cdot \bm{N})^2 \, \dot r^2 -
5(1-3\nu) (\bm{n}\cdot \bm{N})(\bm{v}\cdot \bm{N})\dot r+
\frac{5}{3}(1-3\nu) (\bm{v}\cdot \bm{N})^2\right.\nonumber\\
& \left. \; \; \; + \frac{1}{12} (11-25\nu) v^2-
\frac{3}{4}(1-3\nu) (\bm{n}\cdot \bm{N})^2 \, v^2 -
\frac{G m}{3r} (11+2\nu)-\frac{G m}{r}(1-3\nu) (\bm{n}\cdot \bm{N})^2\right]
\, \nonumber\\
& +  (\bm{n}\times\bm{N})^{a} S_{\rm c}^{ b}
\ \delta\left[-\frac{5}{4}\, \dot r^2+
\frac{15}{4} (\bm{n}\cdot \bm{N})^2 \, \dot r^2-
5(\bm{n}\cdot \bm{N})(\bm{v}\cdot \bm{N}) \, \dot r+
\frac{5}{3}(\bm{v}\cdot \bm{N})^2+\frac{1}{4} \, v^2\right. \nonumber\\
& \left. \; \; \;   -\frac{3}{4} (\bm{n}\cdot \bm{N})^2 \, v^2 -
\frac{G m}{r} (\bm{n}\cdot \bm{N})^2\right] \, +
\, (\bm{n}\times\bm{v})^{a}\Sigma_{\rm c}^{b}
\ (1-4\nu) \bigg[2(\bm{v}\cdot \bm{N})-
2(\bm{n}\cdot \bm{N})  \dot r\bigg] \nonumber\\
& +n^{a} \, v^{ b}\left[36 (-1+4\nu) (\bm{n}\cdot \bm{N})
(\bm{n}\times \bm{v}) \cdot \bm{\Sigma}_{\rm c} -4(2-9\nu) \, \dot r
\, (\bm{n}\times \bm{\Sigma}_{\rm c})\cdot \bm{N}+
\frac{2}{3}(13-55\nu) (\bm{v}\times
\bm{\Sigma}_{\rm c})\cdot \bm{N}
\right.\nonumber\\
&\left. \; \; \; +\delta \left(
 -72 (\bm{n}\cdot \bm{N}) (\bm{n}\times \bm{v}) \cdot \bm{S}_{\rm c}-
20 \, \dot r\, (\bm{n}\times \bm{S}_{\rm c})\cdot \bm{N}+
\frac{50}{3} (\bm{v}\times \bm{S}_{\rm c})\cdot \bm{N}\right)\right]
\nonumber\\
&+ (\bm{n}\times\bm{v})^{a} S_{\rm c}^{b}
\delta\left[-6 (\bm{n}\cdot \bm{N})  \dot r+
\frac{14}{3}(\bm{v}\cdot \bm{N}) \right]\, +
n^{a} (\bm{v}\times \bm{S}_{\rm c})^{b} \delta
\bigg[-26 \, \dot r \, (\bm{n}\cdot \bm{N})+
12 (\bm{v}\cdot \bm{N})\bigg]\nonumber\\
& +  n^{a}(\bm{v}\times \bm{\Sigma}_{\rm c})^{b}\left[
2(-7+29\nu) \, \dot r \, (\bm{n}\cdot \bm{N}) +
\frac{2}{3}(10-43\nu) (\bm{v}\cdot \bm{N})\right] \, +
\, v^{a}(\bm{v}\times\bm{S}_{\rm c})^{b} \,  \delta\,
\frac{64}{3} (\bm{n}\cdot \bm{N})\nonumber\\
& + v^{a}(\bm{n}\times \bm{\Sigma}_{\rm c})^{b}
\left[-2(5-22\nu)\, \dot r \, (\bm{n}\cdot \bm{N})+\frac{4}{3}\left(1-
6 \nu\right) (\bm{v}\cdot \bm{N}) \right]  \, +
\,  v^{a}(\bm{v}\times \bm{\Sigma}_{\rm c})^{b}
\ \frac{2}{3} (16-67 \nu) (\bm{n}\cdot \bm{N})\nonumber\\
&  +v^{a}(\bm{n}\times \bm{S}_{\rm c})^{b} \delta\left[
-26 \, \dot r \, (\bm{n}\cdot \bm{N})+
\frac{4}{3}(\bm{v}\cdot \bm{N})\right]\, +
\, v^{a} (\bm{n}\times \bm{v})^{b} \left[
2(-1+4\nu) (\bm{N}\cdot \bm{\Sigma}_{\rm c}) -
\frac{14}{3}\delta\, (\bm{N}\cdot \bm{S}_{\rm c})\right]
\nonumber\\
& + v^{a}(\bm{n}\times \bm{N})^{b}\left[-(3-23\nu)\,
\dot r\, (\bm{n}\cdot\bm{\Sigma}_{\rm c})-
5 (1-3\nu)\, \dot r\, (\bm{n}\cdot \bm{N})(\bm{N}\cdot \bm{\Sigma}_{\rm c})+
\frac{2}{3}(1+8\nu) (\bm{v}\cdot \bm{\Sigma}_{\rm c})\right.\nonumber\\
&\left. \; \; \;  +
\frac{10}{3}(1-3\nu)(\bm{v}\cdot \bm{N}) (\bm{N}\cdot \bm{\Sigma}_{\rm c})+
\delta \left(
\frac{10}{3} (\bm{v}\cdot \bm{N}) (\bm{N}\cdot \bm{S}_{\rm c})-
11  \, \dot r\, (\bm{n}\cdot \bm{S}_{\rm c})-
5 \, \dot r\, (\bm{n}\cdot \bm{N})(\bm{N}\cdot \bm{S}_{\rm c})
\right.\right.\nonumber\\
& \; \; \;   -
\frac{2}{3} (\bm{v}\cdot \bm{S}_{\rm c}) \bigg)\bigg] \, +
\, S_{\rm c}^{a} (\bm{v}\times \bm{N})^{b} \delta\left[
\frac{5}{6}\, \dot r-\frac{5}{2} \, \dot r \, (\bm{n}\cdot \bm{N})^2+
\frac{10}{3}(\bm{v}\cdot \bm{N})(\bm{n}\cdot \bm{N})\right]\nonumber\\
& + \Sigma_{\rm c}^{a} (\bm{v}\times \bm{N})^{b} \left[
-\frac{29}{6}(1+\nu) \, \dot r -
\frac{5}{2} (1-3\nu) \, \dot r \, (\bm{n}\cdot \bm{N})^2+
\frac{10}{3} (1-3\nu) (\bm{v}\cdot \bm{N}) (\bm{n}\cdot \bm{N}) \right]
\nonumber\\
& + v^{a} (\bm{v}\times \bm{N})^{b}\left[
-\frac{40\nu}{3} (\bm{n}\cdot \bm{\Sigma}_{\rm c})+
\frac{10}{3} (1-3\nu)(\bm{n}\cdot \bm{N}) (\bm{N}\cdot \bm{\Sigma}_{\rm c}) +
\delta \left(\frac{20}{3} (\bm{n}\cdot \bm{S}_{\rm c}) +
\frac{10}{3} (\bm{n}\cdot \bm{N})(\bm{N}\cdot \bm{S}_{\rm c}) \right)\right]
\nonumber\\
&+  v^{a} \, v^{b} \left[
 \left(\frac{2}{3}-4\nu\right)(\bm{n}\times \bm{\Sigma}_{\rm c})\cdot \bm{N}+
\frac{2}{3} \delta\,(\bm{n}\times \bm{S}_{\rm c})\cdot \bm{N}
\right]\nonumber\\
&\, +({\bm\Sigma}_{\rm c} \times
\bm{N})^{a}n^{b}\bigg[\frac{5}{4}(1+7\nu)\dot r^2+\frac{15}{4}(1-3\nu)\dot
r^2(\bm{n}\cdot \bm{N})^2+5(-1+3\nu)\dot r(\bm{n}\cdot \bm{N})(\bm{v}\cdot
\bm{N})+\frac{5}{3}(1-3\nu)(\bm{v}\cdot \bm{N})^2\nonumber\\
& \; \; \; +\frac{1}{12}(11-25\nu)v^2+\frac{3}{4}(-1+3\nu)(\bm{n}\cdot
\bm{N})^2 v^2
+\frac{G m}{3r}(-17+10\nu)+\frac{G m}{r}(-1+3\nu)(\bm{n}\cdot
\bm{N})^2\bigg]\nonumber\\
&  +({\bm S}_{\rm c}\times \bm{N})^{a}{n}^{b}\delta\bigg[-\frac{5}{4}\dot
r^2+\frac{15}{4}\dot r^2({\bm n}\cdot \bm{N})^2-5\dot r ({\bm n}\cdot
\bm{N})({\bm v}\cdot \bm{N})+\frac{5}{3}({\bm v}\cdot
\bm{N})^2+\frac{1}{4}v^2-\frac{3}{4}v^2 ({\bm n}\cdot \bm{N})^2\nonumber\\
& \; \; \; -\frac{2G m}{r}-
\frac{G m}{r}({\bm n}\cdot \bm{N})^2\bigg]\nonumber\\
&+(\bm{\Sigma}_{\rm c}\times \bm{N})^{a}v^{b}\bigg[-\frac{29}{6}(1+\nu)\dot
r+\frac{5}{2}(-1+3\nu)\dot r(\bm{n}\cdot
\bm{N})^2+\frac{10}{3}(1-3\nu)(\bm{n}\cdot \bm{N})(\bm{v}\cdot
\bm{N})\bigg]\nonumber\\
&  +(\bm{S}_{\rm c}\times \bm{N})^{a}v^{b}\delta\bigg[\frac{5}{6}\dot
r-\frac{5}{2}\dot r (\bm{n}\cdot \bm{N})^2+\frac{10}{3}(\bm{n}\cdot
\bm{N})(\bm{v}\cdot \bm{N})\bigg]\nonumber\\
& +(\bm{v}\times \bm{N})^{a}n^{b}\bigg[(-3+23 \nu)\dot r (\bm{n}\cdot
\bm{\Sigma}_{\rm c})+5(-1+3\nu) \dot r(\bm{n}\cdot \bm{N})(\bm{\Sigma}_{\rm
  c}\cdot \bm{N})
+\frac{1}{3}(5+7\nu) (\bm{v}\cdot \bm{\Sigma}_{\rm c})\nonumber\\
& \; \; \; +\frac{10}{3}(1-3\nu)(\bm {v}\cdot \bm{N})(\bm{\Sigma}_{\rm
  c}\cdot \bm{N})+ \delta \left(-11\dot r (\bm{n}\cdot \bm{S}_{\rm c})-5 \dot
  r (\bm{n}\cdot \bm{N})(\bm{S}_{\rm c}\cdot \bm{N})
+\frac{1}{3}(\bm{v}\cdot \bm{S}_{\rm
  c})+\frac{10}{3}(\bm{v}\cdot\bm{N})(\bm{S}_{\rm c}\cdot \bm{N})\right)
\bigg] \ \ \ \ \ \
\end{align}
\end{widetext}
These contributions add linearly to the other known terms in the
waveform. Note that in Eq. (\ref{eq:hijSO}) we have already
anticipated the transverse-traceless projection and simplified the
expression using
$\delta^{ij}_{\mathrm{TT}}=N^i_{\mathrm{TT}}=N^j_{\mathrm{TT}}=0$
and the interchange identity \cite{Kidder:1995zr}
\begin{equation}
 {\cal P}_{ijab}^{\mathrm{TT}} \  A^a (\bm{B}\times \bm{N})^b=
{\cal P}_{ijab}^{\mathrm{TT}}\  B^a(\bm{A}\times \bm{N})^b,
\end{equation}
for any vectors $\bm{A}$ and $\bm{B}$.

\subsection{Spin-spin effects}
\label{sec:SS}

Spin-spin terms in the waveform at 2PN order are entirely
attributable to the equations of motion; they arise when
substituting $\bm{a}^{\rm SS}$ in the time derivatives of
$I_{ab}^{\rm Newt}$. The second time derivative of the contribution
$I_{ab}^{\text{S}^2}$ given in Eq.~\eqref{eq:ISSresult} is at least
of 3PN order (because of the fact that spins are constant at leading
approximation) and therefore vanishes for our calculation. We derive
 \begin{align}
\label{eq:hijSS}
  h^{\rm 2PN  SS}_{{ij \   {\mathrm{TT}}}} &=\frac{6G^2 \nu}{r^3 R} {\cal
    P}_{ijab}^{\mathrm{TT}}
\bigg\{ \nonumber \\ & \quad ~ n^a \, n^b  \Big[5 (\bm{n}\cdot
\bmSeffp)(\bm{n}\cdot \bmSeffm)-
 (\bmSeffp\cdot \bmSeffm) \Big]\nonumber\\
& \qquad ~ -  n^a \, \Seffp^b (\bm{n}\cdot \bmSeffm)- n^a \,
\Seffm^b (\bm{n}\cdot \bmSeffp)  \bigg\}\,.
\end{align}
We notice that the spin-orbit contributions at 2PN order are zero
for an equal-mass, equal-spin black-hole binary. This is a
consequence of the multipoles (\ref{spinmultis}) being zero for this
highly symmetric binary configuration.

The general results (\ref{eq:hijSO}) and (\ref{eq:hijSS}) are
available as a \textsc{mathematica} notebook upon request to be used
to compute the gravitational polarizations and spherical harmonic
modes for precessing binaries for any choice of the source frame and
the polarization triad~\cite{Finn1993,
  Kidder:1995zr,Buonanno:2002fy,Schmidt:2010it,OShaughnessy2011, Ochsner2012,
  2011PhRvD..84l4011B,Schmidt:2012rh}. Below, we shall derive the
polarizations and spin-weighted spherical-harmonic modes for the
case of nonprecessing compact binaries on circular orbits.


\subsection{Reduction to quasicircular orbits}

We now specialize Eqs.~(\ref{eq:hijSO}) and (\ref{eq:hijSS}) to the
case of orbits that have a constant separation $r$ in the absence of
radiation reaction and for which the precession time scale is much
longer than an orbital period.  The details of the derivation of the
modified Kepler law relating the orbit-averaged orbital angular
frequency $\omega$ and the orbit-averaged orbital separation are
discussed in Ref.~\cite{Racine2008}. The instantaneous accelerations
(\ref{eq:eom}) and (\ref{eq:ssaccel}) are projected onto a triad
consisting of the following unit vectors: $\bm{n}=\bm{x}/r$, the
vector $\bm{\ell}=\bm{L}_{\rm N}/|\bm{L}_{\rm N}|$ orthogonal to the
instantaneous orbital plane, where $\bm{L}_{\rm N}=m\nu\,
\bm{x}\times \bm{v}$ denotes the Newtonian orbital angular momentum,
and $\bm{\lambda}=\bm{\ell}\times \bm{n}$. The orbital separation
$r$ and angular frequency $\omega$ are decomposed into their orbit
averaged piece, indicated by an overbar, and remaining fluctuating
pieces, $r=\bar r+\delta r$ and $\omega=\bar \omega+\delta \omega$.
Projecting the equations of motion along $\bm{\lambda}$ yields the
equality $2 \omega \, \dot{r} + \dot{\omega}\, r$ or,
equivalently~\cite{Racine2008}
\begin{equation}
\frac{d}{dt} (\omega\, r^2) = - \frac{3G}{2 m \omega\,r^3c^4} \frac{d}{dt}
(\bm{n} \cdot \bmSeffp) (\bm{n} \cdot \bmSeffm) \, .
\end{equation}
At the 2PN order, $r$ and $\omega$ can be replaced by the constants
$\overline{r}$ and $\overline{\omega}$, respectively, on the right-hand side.
The expression for $\omega\, r^2$ follows from (i) dropping the time
derivatives in the above equation, and (ii) adding an integration constant
determined by averaging $\omega\, r^2$ over an orbit. Inserting the result in
the projection along $\bm{n}$ of the equations of motion,
\begin{equation}
\ddot{r} - \omega^2 r =  (\bm{n}\cdot \bm{a}) \,
\end{equation}
and linearizing in $\delta r$ we find an explicit solution to the differential
equation given by
\begin{subequations}
\begin{align}
\label{eq:rdot}
\dot{r} &= \frac{d\delta r}{dt} \nonumber \\ &
= - \frac{\omega}{2 m^2 r c^4}
[(\bm{n} \cdot \bmSeffp) (\bm{\lambda} \cdot \bmSeffm) +
(\bm{\lambda}\cdot \bmSeffp) (\bm{n}\cdot \bmSeffm)]  \, ,\\
\omega^2 &= \frac{\ddot{r}-(\bm{n} \cdot \bm{a})}{r} \nonumber \\*
&  = \frac{G m}{r^3} \Big[1 - (3-\nu) \frac{G m}{rc^2} \nonumber \\*
& \qquad  \quad- \Big(\frac{G m}{r c^2}\Big)^\frac{1}{2}
\frac{5 (\bm{\ell}\cdot \bm{S}_{\rm c})+ 3 \delta\, (\bm{\ell}\cdot
  \bm{\Sigma}_{\rm c})}{mrc^2}
\nonumber \\* & \qquad \quad + \frac{1}{2m^2 r^2c^4} \Big((\bmSeffp
\cdot
\bmSeffm) +  2 (\bm{\ell}\cdot \bmSeffp) (\bm{\ell}\cdot
\bmSeffm)\nonumber \\*
& \qquad \quad ~-5 (\bm{n}\cdot\bmSeffp)
(\bm{n}\cdot \bmSeffm)\Big) \Big] \, .  \label{eq:omegaofr}
\end{align}
\end{subequations}
Inverting Eq.~(\ref{eq:omegaofr}) to write $r$ as a function of $\omega$ in
Eq. (\ref{eq:hijSO}) and inserting there the expression~\eqref{eq:rdot} of
$\dot{r}$, we obtain the following spin-orbit terms in the waveform:
\begin{widetext}
\begin{eqnarray}
\label{eq:hijSOcirc}
h_{ij\  {\rm{TT}}}^{\mathrm {2PN SO}}&=& \frac{G^2 \nu m \omega^2}{3R }{\cal
  P}_{ijab}^{\mathrm{TT}}\Bigg\{
 n^a \, n^b\, \left[
4 (1-7\nu)(\bm{\ell}\cdot \bm{\Sigma}_{\rm c})(\bm{\lambda}\cdot \bm{N})-
(13-59\nu)(\bm{n}\times \bm{\Sigma}_{\rm c})\cdot \bm{N}-
21\delta\,(\bm{n}\times \bm{S}_{\rm c})\cdot \bm{N}\right]\nonumber\\
 && \, + \lambda^a \, \lambda^b \, \left[
4(7-24\nu)(\bm{\ell}\cdot \bm{\Sigma}^{\rm c})(\bm{\lambda}\cdot \bm{N})+
4(1-6\nu)(\bm{n}\times \bm{\Sigma}^{\rm c})\cdot \bm{N}+
\delta  \bigg(4 (\bm{n}\times \bm{S}^{\rm c})\cdot \bm{N}+
52 ( \bm{\ell}\cdot \bm{S}^{\rm c})(\bm{\lambda}\cdot \bm{N})\bigg)\right]
\nonumber\\
&& \, + \lambda^{a} \, n^{b} \, \bigg[
4(13-55\nu)(\bm{\lambda}\times \bm{\Sigma}_{\rm c}) \cdot \bm{N}+
2(-63+239\nu)(\bm{n}\cdot \bm{N})(\bm{\ell}\cdot \bm{\Sigma}_{\rm c})
\nonumber\\
&& \left. \; \; \; \; \;  +
\delta \bigg(100  (\bm{\lambda}\times \bm{S}_{\rm c})\cdot \bm{N}-
262(\bm{n}\cdot \bm{N})(\bm{\ell}\cdot \bm{S}_{\rm c}) \bigg)\right] \,
+ \Sigma_{\rm c}^{a}\,  \ell^{b} \, 12 (1-4\nu) (\bm{\lambda}\cdot \bm{N})
\nonumber\\
&&+ \lambda^{a} \, {\ell}^{b}\bigg[
12(-1+4\nu)(\bm{N} \cdot \bm{\Sigma}_{\rm c})+
8(1-6\nu) (\bm{\lambda}\cdot \bm{\Sigma}_{\rm c}) (\bm{\lambda}\cdot \bm{N})+
4 (-16+67\nu) (\bm{n}\cdot \bm{\Sigma}_{\rm c}) (\bm{n}\cdot \bm{N}) \nonumber\\
&& \left. \; \; \; \; \; +\delta \bigg(
- 28(\bm{N}\cdot \bm{S}_{\rm c})+
8   (\bm{\lambda}\cdot \bm{S}_{\rm c}) (\bm{\lambda}\cdot \bm{N})-
128(\bm{n}\cdot \bm{S}_{\rm c}) (\bm{n}\cdot \bm{N})\bigg)
 \right]\nonumber\\
&&+ n^{a} \, {\ell}^{b}\bigg[
2(-13+59\nu) (\bm{\lambda}\cdot \bm{\Sigma}_{\rm c})(\bm{n}\cdot \bm{N})+
4(-10+43\nu) (\bm{n}\cdot \bm{\Sigma}_{\rm c})(\bm{\lambda}\cdot \bm{N})
\nonumber\\
&& \left. \; \; \; \; \; +
\delta \bigg(
-42 (\bm{\lambda}\cdot \bm{S}_{\rm c})(\bm{n}\cdot \bm{N})-
72 (\bm{n}\cdot \bm{S}_{\rm c})(\bm{\lambda}\cdot \bm{N})\bigg)\right]\, +
S_{\rm c}^{a}\,  \ell^{b} \, 28\delta\, (\bm{\lambda}\cdot \bm{N})
\nonumber\\
&&+ n^{a}(\bm{n}\times \bm{N})^{b}\left[
-(1+\nu) (\bm{n}\cdot \bm{\Sigma}_{\rm c})-
21(1-3\nu) (\bm{n}\cdot \bm{N}) (\bm{N}\cdot \bm{\Sigma}_{\rm c}) +
\delta \bigg(3(\bm{n} \cdot \bm{S}_{\rm c}) -
21 (\bm{n}\cdot \bm{N}) (\bm{N}\cdot \bm{S}_{\rm c}) \bigg)\right]\nonumber\\
&&+ \lambda^{a}(\bm{n}\times \bm{N})^{b}\left[
2(7+23\nu) (\bm{\lambda}\cdot \bm{\Sigma}_{\rm c})+
40 (1-3\nu)(\bm{N}\cdot \bm{\Sigma}_{\rm c}) (\bm{\lambda} \cdot \bm{N}) +
\delta \bigg(
40 (\bm{N}\cdot \bm{S}_{\rm c}) (\bm{\lambda} \cdot \bm{N})  -
2(\bm{\lambda}\cdot \bm{S}_{\rm c})\bigg)\right]\nonumber\\
&& + \Sigma_{\rm c}^{a} (\bm{n}\times \bm{N})^{b}\bigg[-(21+17\nu)+
20 (1-3\nu) (\bm{\lambda}\cdot \bm{N})^2+
21(-1+3\nu)(\bm{n}\cdot \bm{N})^2\bigg]\nonumber\\
&&+ S_{\rm c}^{a} (\bm{n}\times \bm{N})^{b} \delta \bigg[
-9+20 (\bm{\lambda}\cdot \bm{N})^2-21(\bm{n}\cdot \bm{N})^2\bigg]\, +
S_{\rm c}^{a} \, (\bm{\lambda}\times \bm{N})^{b} \, 40
\, \delta \, (\bm{\lambda}\cdot \bm{N}) (\bm{n} \cdot \bm{N})
\nonumber\\
&&+ \lambda^{a} \, (\bm{\lambda}\times \bm{N})^{b} \left[
-80 \nu (\bm{n}\cdot \bm{\Sigma}_{\rm c})+
20(1-3\nu) (\bm{n}\cdot \bm{N}) (\bm{N}\cdot \bm{\Sigma}_{\rm c})+
\delta \bigg(40 (\bm{n}\cdot \bm{S}_{\rm c})+
20 (\bm{N} \cdot \bm{S}_{\rm c}) (\bm{n}\cdot \bm{N})\bigg)\right]\nonumber\\
&&+\Sigma_{\rm c}^{a} \, (\bm{\lambda}\times \bm{N})^{b}
\, 40(1-3\nu) (\bm{\lambda}\cdot \bm{N}) (\bm{n} \cdot \bm{N}) \Bigg\} \,.
\end{eqnarray}
\end{widetext}
Here, we have used that
\begin{equation}
 (\bm{n}\times\bm{S}_{\rm c})^i=-\lambda^i(\bm{\ell}\cdot \bm{S}_{\rm c})+
\ell^i(\bm{\lambda}\cdot \bm{S}_{\rm c}),
\end{equation}
and similarly for $\bm{\Sigma}_{\rm c}$.

Finally, we derive the 2PN spin-spin terms for circular orbits. They read
\begin{align}
& h_{ij \ \mathrm{ TT}}^{\mathrm {2 PN SS}}= \frac{2G  \nu \omega^2}{m R} {\cal
  P}_{ijab}^{\mathrm{TT}} \bigg\{
n^a\, n^b
 \Big[-\frac{8}{3} ( \bmSeffp\cdot \bmSeffm)
\nonumber \\ & \qquad  \qquad \quad +
\frac{2}{3}
(\bm{\ell}\cdot \bmSeffp) (\bm{\ell}\cdot
\bmSeffm) +
\frac{40}{3} (\bm{n}\cdot \bmSeffp) (\bm{n}\cdot
\bmSeffm) \Big]\nonumber \\ & \qquad +
 \lambda^a \, \lambda^b
\Big[ \frac{2}{3} ( \bmSeffp\cdot \bmSeffm) + \frac{4}{3}
(\bm{\ell}\cdot \bmSeffp) (\bm{\ell}\cdot \bmSeffm)
\nonumber \\ & \qquad \qquad \quad -
\frac{10}{3} (\bm{n}\cdot \bmSeffp) (\bm{n}\cdot \bmSeffm) \Big]
\nonumber \\ & \qquad -2
n^a \, \lambda^b \Big[ (\bm{n}\cdot \bmSeffp) (\bm{\lambda}\cdot
\bmSeffm)+
(\bm{n}\cdot \bmSeffm) (\bm{\lambda}\cdot \bmSeffp)\Big]
\nonumber \\ & \qquad -3
(\bm{n}\cdot \bmSeffp)\,  n^{(a} \, \Seffm^{b)}
- 3 (\bm{n}\cdot \bmSeffm) \, n^{(a} \, \Seffp^{b)}
\bigg\} \, .
\end{align}
%

\subsection{Polarizations for nonprecessing, spinning compact bodies}
\label{sec:pol}

The two polarization states $h_+$ and $h_\times$ are obtained by choosing a
coordinate system and taking linear combinations of the components of
$h_{ij}^\mathrm{TT}$. Using an orthonormal triad consisting of $\bm{N}$ and two
polarization vectors $\bm{P}$ and $\bm{Q}$, the polarizations are
\begin{subequations}
 \label{eq:polarizations}
\begin{eqnarray}
h_+&=& \frac{1}{2}\left(P^iP^j-Q^iQ^j\right)h_{ij}^\mathrm{TT}\,,\\
h_\times&=&\frac{1}{2}\left(P^iQ^j+Q^iP^j\right)h_{ij}^\mathrm{TT}\, .
\end{eqnarray}
\end{subequations}
Although different choices of $\bm{P}$ and $\bm{Q}$ give different
polarizations, the particular linear combination of $h_+$ and
$h_\times$ corresponding to the physical strain measured in a
detector is independent of the convention used. For nonspinning
binaries, one usually chooses a coordinate system such that the
orbital plane lies in the $x\mbox{-}y$ plane, and the direction of
gravitational-wave propagation $\bm{N}$ is in the $x\mbox{-}z$
plane.

When the spins of the bodies are aligned or anti-aligned with the
orbital angular momentum, the system's evolution is qualitatively
similar to the case of nonspinning bodies. This case is characterized
by the absence of precession of the spins and orbital angular momentum
and thus the orbital plane remains fixed in space. However, the effect
of the spins gives a contribution to the phase and a correction to the
amplitude of the waveform, which we explicitly provide in this
subsection.  We use the conventions that the $z$ axis coincides with
$\bm{\ell}$ and the vectors $ \bm{\ell}$, $\bm{N}$, $\bm{n}$, and $\bm{\lambda}$ have
the following $(x,y,z)$ components:
\begin{subequations}
\begin{eqnarray}
\bm{\ell}&=& (0,0,1), \; \; \; \; \;  \; \; \; \; \;  \; \; \; \; \; \; \; \; \; \bm{N}=
(\sin\theta, 0 ,\cos\theta), \; \; \; \; \; \; \; \ \; \; \; \; \; \\
 \bm{n}&=& (\sin\Phi, -\cos\Phi, 0), \ \ \ \  \bm{\lambda}=
(\cos \Phi, \sin \Phi, 0),  \ \ \ \; \; \; \; \; \; \; \; 
\end{eqnarray}
\end{subequations}
where $\Phi$ is the orbital phase defined such that at the initial time,
$\bm{n}$ points in the $x$ direction. We use the following polarization
vectors:
\begin{equation}
\bm{P}=\bm{N}\times \bm{\ell}, \ \ \ \bm{Q}=\bm{N}\times \bm{P}.
\end{equation}
The vector $\bm{P}$ is the ascending node where the orbital
separation vector crosses the plane of the sky from below. With
these conventions, Eqs. (\ref{eq:polarizations}) with Eqs.
(\ref{eq:hijSOcirc}), specialized to the case where the only
nonvanishing spin components are $(\bm{\Sigma^{\rm c}}\cdot
\bm{\ell})$ and $(\bm{S}^{\mathrm{c}}\cdot\bm{\ell})$, become
\begin{widetext}
\begin{eqnarray}
\label{eq:hplus}
h_+^{\mathrm{2PN \ spin} }
&=&-\frac{G^2\nu m \omega^2}{12 R }\cos\Phi \ \sin\theta\left\{
3 \delta\, (\bm{\ell}\cdot \bm{S}_{\rm c})(-33+\cos ^2\theta)+
\left[(-93+167\nu)+
9(1-3\nu)\cos^2\theta\right]( \bm{\ell}\cdot \bm{\Sigma_{\rm c}})\right\}
\nonumber\\
&&-\frac{9G^2\nu m\omega^2}{4 R  }\cos (3\Phi) \ \sin\theta \left\{
\delta\,(5-\cos^2\theta)( \bm{\ell}\cdot \bm{S}_{\rm c}) +
3(1-3\nu)\sin^2\theta (\bm{\ell}\cdot \bm{\Sigma_{\rm c}})\right\}\nonumber \\
&&-\frac{2G \nu \omega^2}{m R }\cos (2\Phi)\left(1+\cos^2 \theta\right)
( \bm{\ell}\cdot \bmSeffp )(\bm{\ell}\cdot \bmSeffm )\,
,\ \\
h_\times^{\mathrm{2PN \ spin} }&=&
-\frac{G^2\nu m  \omega^2}{48 R  }\sin\Phi \sin(2\theta)\left\{
6\delta\, ( \bm{\ell}\cdot \bm{S}_{\rm c})
\left(-33+\cos^2\theta\right)+ \left[(-171+289 \nu)+
3(1-3\nu)\cos(2\theta)\right]( \bm{\ell}\cdot \bm{\Sigma}_{\rm c})\right\}
\nonumber\\
&&-\frac{9G^2\nu m \omega^2}{8 R }\sin(3\Phi)\sin(2\theta)\left\{
\delta\, ( \bm{\ell}\cdot \bm{S}_{\rm c})\left(7-3\cos^2\theta\right)+
3(1-3\nu)\sin^2\theta ( \bm{\ell} \cdot \bm{\Sigma_{\rm c}})\right\}\nonumber\\
&&-\frac{4G\nu \omega^2}{m R}\sin(2\Phi)\cos\theta
( \bm{\ell}\cdot \bmSeffp )( \bm{\ell}\cdot \bmSeffm
)\,. \label{eq:hcross}
\end{eqnarray}
\end{widetext}
Here, the convention for the 2PN spin pieces of the polarizations is analogous to that adopted for the PN expansion of the waveform  (\ref{eq:hexpansion}), with the expansion coefficients related by Eqs. (\ref{eq:polarizations}) at each PN order.

\subsection{Gravitational modes for nonprecessing, spinning compact bodies}
\label{sec:modes}
The gravitational wave modes are obtained by expanding the complex polarization
\begin{equation}
h =  h_+
- i h_\times\,,
\label{hcomplex}
\end{equation}
into  spin-weighted $s=-2$ spherical harmonics as
\begin{equation}\label{eq:modeexp}
h(\theta,\phi) = \sum_{\ell = 2}^{+\infty} \sum_{m=-\ell}^{\ell} h_{\ell m}\,
{}_{-2}Y^{\ell m}(\theta,\phi) \, ,
\end{equation}
where
\begin{equation}
{}_{-s} Y^{\ell m}(\theta,\phi) = (-1)^s \sqrt{\frac{2\ell + 1}{4\pi}}
\, d_{sm}^\ell(\theta)\, e^{i m \phi}\,,
\end{equation}
with
\begin{eqnarray}
&&d_{sm}^\ell(\theta) = \sum_{k=\max(0,m-s)}^{\min(\ell+m,\ell-s)}
\frac{(-1)^k}{k!} \nonumber\\
&&\times
\frac{\sqrt{(\ell + m)! (\ell - m)!(\ell + s)!(\ell - s)!}}{(k - m + s)!
    (\ell + m - k)!  (\ell - k - s)!} \ \ \ \ \ \ \ \ \ \nonumber\\
&&\times \left(\cos (\theta/2)\right)^{2\ell+m-2k-s}
\left( \sin (\theta/2) \right)^{2k-m+s} \,.
\end{eqnarray}
The modes $h_{\ell m}$ can be extracted by computing
\begin{equation}
\label{eq:hlm}
h_{\ell m} = \int d\Omega  \, h(\theta, \phi)
\, {}_{-2}{Y}^{\ell  m*}(\theta,\phi) \,,
\end{equation}
where the integration is over the solid angle $\int d\Omega=\int^{\pi}_0
\sin\theta d\theta \int^{2\pi}_0d\phi $
and using the orthogonality property
$ \int d\Omega   \ {}_{-s}Y^{\ell m}(\theta,\phi)
\, {}_{-s}Y^{\ell' m' *}(\theta,\phi) = \delta^{\ell \ell'} \delta^{m m'}$,
where $ \delta^{\ell \ell'}$ is the Kronecker symbol and the star denotes
complex conjugation.
Using Eqs. (\ref{eq:hplus}) and (\ref{eq:hcross}) in Eq.
(\ref{eq:hlm}) we find the following nonvanishing modes:
\begin{equation}
( h_{\ell m})^\mathrm{2PN\,spin }=-\frac{2 G^2 m \nu \,
  \omega^2}{R}\sqrt{\frac{16\pi}{5}}e^{-im\Phi}
\ \hat h_{\ell m},
\end{equation}
\begin{subequations}
\label{eq:hlmnonprec}
\begin{eqnarray}
\hat h_{21}&=&
 -\frac{43}{21} \delta\, (\bm{\ell}\cdot \bm{S}_{\rm c})+
\frac{1}{42}(-79+139\nu)(\bm{\ell}\cdot \bm{\Sigma}_{\rm c})\,, \nonumber \\\\
\hat h_{22}&=& \frac{(\bm{\ell}\cdot
\bmSeffp)(\bm{\ell}\cdot \bmSeffm )}{G m^2}
\,, \label{eq:h22} \\
\hat h_{31}&=& \frac{1}{24\sqrt{14}} \delta\,
(\bm{\ell}\cdot \bm{S}_{\rm c})+
\frac{5}{24\sqrt{14}}(1-3\nu)(\bm{\ell}\cdot \bm{\Sigma}_{\rm c})
\,,\nonumber \\ \\
\hat h_{33}&=&-\frac{3\sqrt{105}}{8\sqrt{2}}
\delta\, (\bm{\ell}\cdot \bm{S}_{\rm c})-
\frac{9}{8}\sqrt{\frac{15}{14}}(1-3\nu)(\bm{\ell}\cdot \bm{\Sigma_{\rm c}})\,,
\nonumber \\ \\
\hat h_{41}&=&\frac{\sqrt{5} }{168\sqrt{2}}\delta\,
(\bm{\ell}\cdot \bm{S}_{\rm c})+
\frac{\sqrt{5} }{168\sqrt{2}}(1-3\nu)(\bm{\ell}\cdot \bm{\Sigma}_{\rm c})\,,
\nonumber \\\\
\hat h_{43}&=&\frac{9\sqrt{5}}{8\sqrt{14}}\delta\,
(\bm{\ell}\cdot \bm{S}_{\rm c})+
\frac{9\sqrt{5}}{8\sqrt{14}}(1-3\nu)(\bm{\ell}\cdot \bm{\Sigma}_{\rm c})
\,.\ \nonumber \\
\end{eqnarray}
\end{subequations}
We have explicitly checked that in the test-mass limit $\nu\to 0$, Eqs.
(\ref{eq:hlmnonprec}) reduce to the 2PN ${\cal O}(q)$ and ${\cal O}(q^2)$
terms given in Eqs. (22) of Ref.~\cite{Pan2010hz} (see also \cite{Tagoshi:1996gh}), after
accounting for the factor of $(-i)^m$ attributable to the different conventions for the
phase origin, as explained in Ref.~\cite{Arun:2009}.

It is interesting to note from Eq. (\ref{eq:h22}) that in the
nonprecessing case, the dominant $h_{22}$ mode contains only terms
that are quadratic in the spin at 2PN order. By contrast, for
precessing binaries, the 2PN spin-orbit terms will give a
nonvanishing contribution to the $22$-mode.

\section{CONCLUSIONS}
\label{sec:conclusions}

We have extended the knowledge of the spin terms in the gravitational-wave
strain tensor to 2PN accuracy for precessing binaries. Our result includes the
spin-orbit as well as the spin${}_1$-spin${}_2$ and spin${}^2_1$,
spin${}_{2}^2$ effects. The quadratic-in-spin terms are entirely due to the
equations of motion, whereas the 2PN spin-orbit terms come from both the
corrections to the orbital dynamics and the radiation field.

For a given choice of an orthonormal polarization triad and a source frame,
the gravitational-wave polarizations can be obtained by projecting our result
for the gravitational-wave strain tensor given in Secs.~\ref{sec:SO}
and~\ref{sec:SS} orthogonal to the propagation direction. For precessing
binaries, there is no preferred unique choice of the source
frame~\cite{Finn1993, Kidder:1995zr,Buonanno:2002fy,Schmidt:2010it,
  OShaughnessy2011, Ochsner2012, 2011PhRvD..84l4011B,Schmidt:2012rh}, but in
the case that the spins are collinear with the orbital angular
momentum, the procedure to obtain the polarizations can be carried
out in a similar fashion as for nonspinning binaries. For the
nonprecessing case and circular orbits, we provided ready-to-use
expressions for the gravitational polarizations in
Sec.~\ref{sec:pol}, which could be directly employed in time-domain
post-Newtonian, phenomenological and effective-one-body--based
template models~\cite{Kidder:1995zr,
Arun:2009,Ajith:2008,Damour2009a,Pan:2009wj,
  Santamaria:2010yb,Pan:2011gk}.

In view of the current interest in interfacing analytical and
numerical relativity, we also provided the decomposition of the
waveform into spin-weighted spherical harmonic modes for
nonprecessing binaries and quasicircular orbits. We verified that
the test-particle limit of our result reduces to the expressions
obtained from black-hole perturbation
theory~~\cite{Tagoshi:1996gh,Pan2010hz}. We noted that for spins
collinear with the orbital angular momentum, the dominant $h_{22}$
mode of the waveform contains only quadratic-in-spin effects since
the spin-orbit contributions vanish in this case, although they are
nonzero for generic, precessing configurations.

\begin{acknowledgments} A.B. acknowledges partial support from NSF Grants
No. PHY-0903631 and No. PHY-1208881, and NASA Grant NNX09AI81G. A.B.
also thanks the Kavli Institute for Theoretical Physics (supported
by the NSF Grant No. PHY11-25915) for hospitality during the
preparation of this manuscript. T.H. acknowledges support from  NSF
Grants No. PHY-0903631 and No. PHY-1208881,  and the Maryland Center
for Fundamental Physics. We thank Gilles Esposito-Far\`ese, Larry
Kidder and Etienne Racine for useful interactions, as well as David
Delavaquerie for help in finalizing one of our \textsc{mathematica}
codes.
\end{acknowledgments}

\appendix*

\section{USEFUL IDENTITIES}

According to the way the waveform is computed, the result may take various
forms, which are not immediately seen to be equivalent. Their difference
vanishes because of some dimensional identities valid in three dimensions. They all
amount to expressing the fact that a tensor with four antisymmetrized indices
must vanish. We shall present here two of such identities, which turned out to
be particularly useful for our checks, together with Eqs.~(5.2) of
Ref.~\cite{Faye-Blanchet-Buonanno:2006}.

Let $\bm{U}_A=U_A^i$, for $A\in \{1,2,3\}$, be three vectors of $\mathbb{R}^3$.
The first identity tells us that for any vector $\bm{U}$, we must have
\begin{align}
& (\bm{U}_1 \times \bm{U}_2)^{(i} [U_3^{j)} (\bm{U}_4 \cdot \bm{U}) - U_4^{j)}
(\bm{U}_3 \cdot \bm{U})]
 \\ & ~ =
U_4^{(i} [(\bm{U}\times \bm{U}_1)^{j)} (\bm{U}_2 \cdot \bm{U}_3) -  (\bm{U}\times
\bm{U}_2)^{j)} (\bm{U}_1 \cdot \bm{U}_3)] \nonumber \\ & \quad + U_3^{(i}
[(\bm{U}\times \bm{U}_2)^{j)} (\bm{U}_1 \cdot \bm{U}_4) - (\bm{U}\times
\bm{U}_1)^{j)} (\bm{U}_2 \cdot \bm{U}_4)] \, .
\nonumber \end{align}
To show this, we compute $\varepsilon^i_{~ab} \varepsilon^{mjk}
\varepsilon_{mpq} U_1^a U_2^b U_3^p U_4^q$ in two different manners:
(i) we group the first two epsilons, which are next expanded in
terms of the identity tensor $\delta^i_{~j}$ using the standard
formula $\varepsilon_{iab} \varepsilon^{mjk} =3! \delta_{~[i}^m
\delta_{~a}^j \delta_{~b]}^k$; (ii) we group the last two epsilons
and apply the contracted version of the previous equation:
$\varepsilon^{mjk} \varepsilon_{mpq} = 2 \delta^j_{~[p}
\delta^k_{~q]}$. One of the remaining free indices, say $k$, is
finally contracted with $U_k$.

The second identity reads:
\begin{widetext}
\begin{align} \label{eq:identity2}
  & \delta^{ij} [U_1^2 U_2^2 U_3^2 -
  U_1^2 (\bm{U}_2 \cdot \bm{U}_3)^2 -
  U_2^2 (\bm{U}_3  \cdot \bm{U}_1)^2 -
  U_3^2 (\bm{U}_1 \cdot \bm{U}_2)^2 + 2 (\bm{U}_1 \cdot \bm{U}_2) (\bm{U}_2
  \cdot \bm{U}_3) (\bm{U}_3 \cdot \bm{U}_1)]
  \nonumber \\ & \qquad +
  2 U_1^{(i} U_3^{j)} [ U_2^2 (\bm{U}_3 \cdot \bm{U}_1) - (\bm{U}_1 \cdot
  \bm{U}_2) (\bm{U}_2 \cdot \bm{U}_3)] + 2 U_1^{(i}
  U_2^{j)} [ U_3^2 (\bm{U}_1 \cdot \bm{U}_2) - (\bm{U}_2 \cdot \bm{U}_3)
  (\bm{U}_3 \cdot \bm{U}_1)]
  \nonumber \\ & \qquad + 2 U_2^{(i} U_3^{j)} [ U_1^2 (\bm{U}_2 \cdot
  \bm{U}_3) - (\bm{U}_1 \cdot \bm{U}_2) (\bm{U}_1 \cdot \bm{U}_3)] +
  U_1^i U_1^j [ (\bm{U}_2 \cdot \bm{U}_3)^2 - U_2^2 U_3^2]
  + U_2^i U_2^j [ (\bm{U}_1 \cdot \bm{U}_3)^2 - U_1^2 U_3^2]
  \nonumber \\ & \qquad +
  U_3^i U_3^j [ (\bm{U}_1 \cdot \bm{U}_2)^2 - U_1^2 U_2^2] =0 \, .
\end{align}
\end{widetext}
It is proved by contracting the equality $U_1^{[a} U_2^b U_3^c
\delta^{i]j}=0$ with $U_{1a} U_{2b} U_{3c}$ and expanding. As the
trace of the left-hand side of Eq.~\eqref{eq:identity2} is
identically zero, the nontrivial content of this identity consists
of its STF part.

\end{document}